\documentclass[aps,prl,twocolumn,amsmath,amssymb,superscriptaddress]{revtex4-1}

\usepackage{stackengine,graphicx,xcolor}
\usepackage{amsmath}
\usepackage{etaremune}
\usepackage{braket}
\usepackage{multirow}
\usepackage{mathtools}
\usepackage{textcomp}
\begin{document}
\title{Absorption-Based Diamond Spin Microscopy on a Plasmonic Quantum Metasurface}
\author{Laura Kim}
\affiliation{Research Laboratory of Electronics, MIT, Cambridge, MA 02139, USA}

\author{Hyeongrak Choi}
\affiliation{Research Laboratory of Electronics, MIT, Cambridge, MA 02139, USA}
\affiliation{Department of Electrical Engineering and Computer Science, MIT, Cambridge, MA 02139, USA}

\author{Matthew E. Trusheim}
\affiliation{Research Laboratory of Electronics, MIT, Cambridge, MA 02139, USA}
\affiliation{U.S. Army Research Laboratory, Sensors and Electron Devices Directorate, Adelphi, Maryland 20783, USA}

\author{Dirk R. Englund}
\affiliation{Research Laboratory of Electronics, MIT, Cambridge, MA 02139, USA}
\affiliation{Department of Electrical Engineering and Computer Science, MIT, Cambridge, MA 02139, USA}
\date{\today}

\begin{abstract}
Nitrogen vacancy (NV) centers in diamond have emerged as a leading quantum sensor platform, combining exceptional sensitivity with nanoscale spatial resolution by optically detected magnetic resonance (ODMR).  Because fluorescence-based ODMR techniques are limited by low photon collection efficiency and modulation contrast, there has been growing interest in infrared (IR)-absorption-based readout of the NV singlet state transition. IR readout can improve contrast and collection efficiency \cite{Acosta2010-rp,Bougas2018-ej,Dumeige2013-wf}, but it has thus far been limited to long-pathlength geometries in bulk samples due to the small absorption cross section of the NV singlet state. Here, we amplify the IR absorption by introducing a resonant diamond metallodielectric metasurface that achieves a quality factor of Q $\sim$ 1,000. This "plasmonic quantum sensing metasurface" (PQSM) combines localized surface plasmon polariton resonances with long-range Rayleigh-Wood anomaly modes and achieves desired balance between field localization and sensing volume to optimize spin readout sensitivity.  From combined electromagnetic and rate-equation modeling, we estimate a sensitivity below 1 nT Hz$^{-\frac{1}{2}}$ per \textmu m$^2$ of sensing area using numbers for present-day NV diamond samples and fabrication techniques. The proposed PQSM enables a new form of microscopic ODMR sensing with infrared readout near the spin-projection-noise-limited sensitivity, making it appealing for the most demanding applications such as imaging through scattering tissues and spatially-resolved chemical NMR detection. 
\end{abstract}

\maketitle

\section{Introduction}

The ability to optically measure quantities such as electric field, magnetic field, temperature, and strain under ambient conditions makes the NV system appealing for a range of wide-field sensing applications, from imaging biological systems \cite{Barry2016-rc} and electrical activity in integrated circuits \cite{Turner2020-rq} to studying quantum magnetism and superconductivity in quantum materials \cite{Thiel2016-yf,Tetienne2015-uv,Du2017-xr,Ku2020-ev}. NV-based magnetometers have shown exceptional sensitivity at room temperature, but conventional fluorescence-based readout methods result in sensitivity values far from the spin projection noise limit primarily due to background fluorescence, poor photon collection efficiency, and low spin-state contrast \cite{Barry2020-ec}. These limitations can be overcome by probing the infrared singlet transition near 1042 nm by absorption. However, this absorption-based readout has only been demonstrated for bulk diamond samples with a large optical path length of millimeters \cite{Dumeige2013-wf,Chatzidrosos2017-vf} to centimeters \cite{Bougas2018-ej,Jensen2014-xu} due to the small absorption cross section of the singlet state transition. This long-pathlength requirement presents the central challenge in IR readout to imaging microscopy, where the sensing depth should commonly be below the micron-scale. The plasmonic quantum sensing metasurface (PQSM) solves this problem by confining vertically incident IR probe light in a few-micron-thick diamond layer with a quality factor near 1,000. The PQSM consists of a metallodielectric grating that couples surface plasmon polariton (SPP) excitations and Rayleigh-Wood anomaly (RWA) modes \cite{Gao2009-fk,Steele2003-nj}. The localized plasmonic resonance causes local field concentration as well as a wavelength-scale field enhancement when coupled with the RWA modes. Unlike fluorescence, the directional reflection (or transmission) can be captured with near-unity efficiency. In particular,  detection of the reflected coherent probe light with a standard camera enables shot-noise limited detection, eliminating the need of single photon detectors. Taken together, our analysis predicts that the PQSM coupled to NV sensing layers can enable a sensitivity below 1 nT Hz$^{-\frac{1}{2}}$ per \textmu m$^2$ of sensing area.

\begin{figure*}
    \centering
    \includegraphics[width=0.95\textwidth]{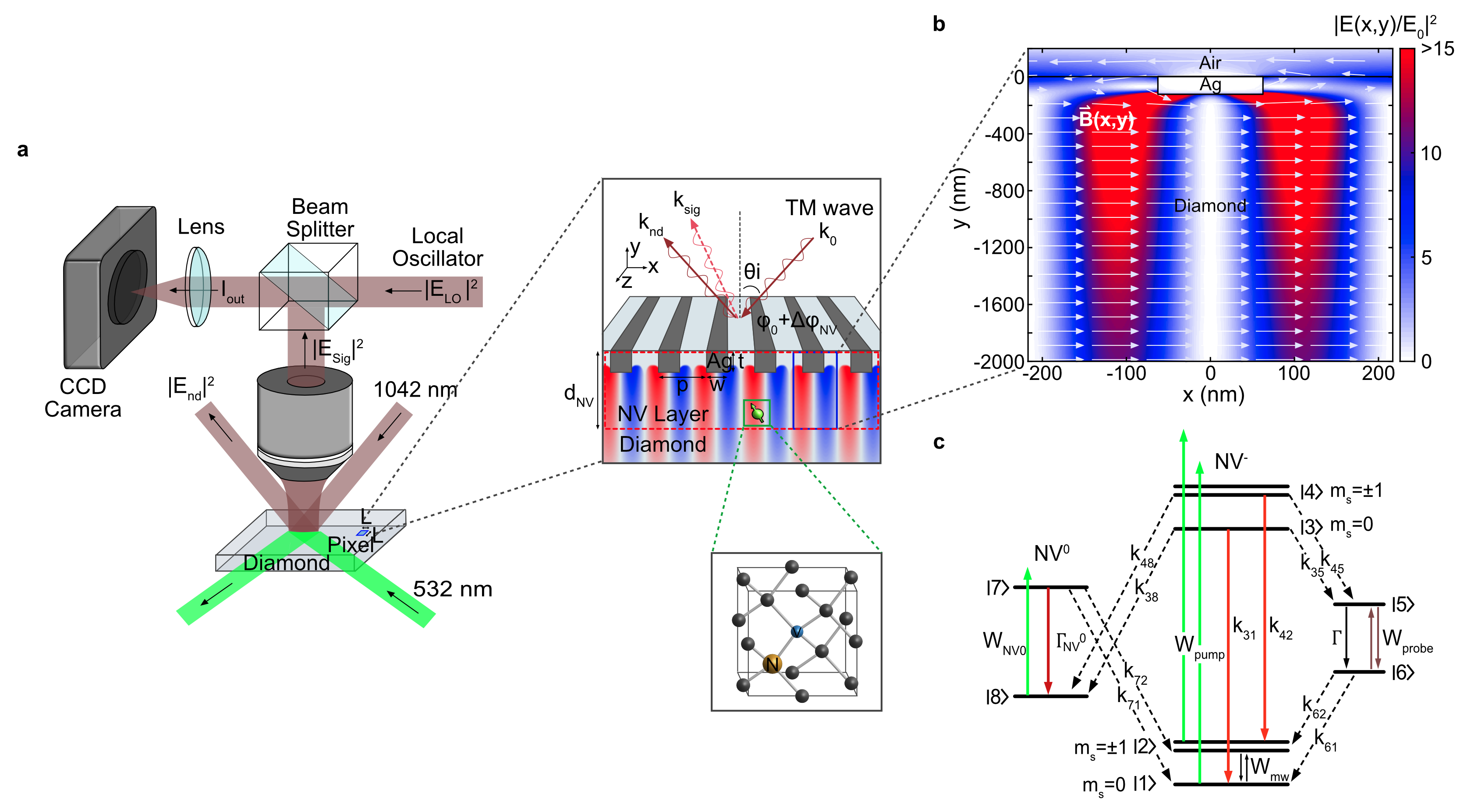}
    \caption{(a) Plasmonic quantum sensing metasurface (PQSM) consisting of a metallodielectric grating and proposed homodyne-detection-based sensing scheme. TM-polarized incoming light induces a SPP-RWA hybrid mode, creating a vertically extended field profile as shown in the the overlapping Re($E_y$). With an applied microwave magnetic field, the PQSM generates spin-dependent reflection with an additional phase change, $\Delta\phi_\text{NV}$. The spin-dependent signal, $|E_\text{sig}|^2$, is separated from the un-diffracted beam, $|E_\text{ud}|^2$, in a dark-field excitation geometry and interferes with a local oscillator, $|E_\text{LO}|^2$. The interfered output beam, $I_\text{out}(I_t,\Omega_R,R,\Delta\phi_\text{LO})$, is detected by a CCD camera.  (c) Total electric field intensity of RWA-SPP resonances upon normal incidence at $\lambda_0$ = 1042 nm with $p$ = 434 nm, $w$ = 125 nm, and $t$ = 125 nm. The arrow plot shows the magnetic field generated by a uniform driving current in an infinite array of plasmonic Ag wires.}
    \label{fig:fig1}
\end{figure*}

\section{Results}
\subsection{IR-absorption-based detection scheme}
The principle of NV-based magnetometers lies in the Zeeman energy shift of the NV defect spin sublevels that can be polarized and measured optically. As illustrated in Fig.\,\ref{fig:fig1}c, the spin sublevels m$_\text{s}$ = 0 and m$_\text{s}$ = $\pm1$ of the $^3$A$_2$ ground triplet state, labelled as $\ket{1}$ and $\ket{2}$, respectively, are separated by a zero-field splitting of D = 2.87 GHz, whose transition can be accessed with a resonant microwave field. Upon spin-conserving off-resonant green laser excitation, a fraction of the population decays non-radiatively into the $^1$A$_1$ metastable singlet state, $\ket{5}$, predominantly from $\ket{4}$ as k$_{45}$ $\gg$ k$_{35}$, where k$_\text{ij}$ indicates the decay rate from level i to level j. After a sub-ns decay from $\ket{5}$ to $\ket{6}$, the shelving time at $\ket{6}$ exceeds 200 ns at room temperature \cite{Acosta2010-cp}. Therefore, the population of $\ket{6}$ can be measured by absorption of the singlet state transition resonant at 1042 nm. \par
Figure\,\ref{fig:fig1}a shows the proposed PQSM for the IR-absorption-based detection scheme. The sensing surface is pumped with a green laser at 532 nm for NV spin  initialization and illuminated with transverse magnetic (TM) polarized probe light at $\lambda_0$ = 1042 nm for IR readout. The PQSM-NV layer causes a spin-dependent phase and amplitude of the IR reflection. The spatially well-defined signal beam is separated from the incident probe field in a dark-field excitation geometry (i.e., k-vector filtering). By interfering with a local oscillator,  phase-sensitive homodyne detection at the camera enables measurement at the photon shot noise limit. The PQSM doubles as a wire array for NV microwave control \cite{Shalaginov2020-sd,Ibrahim2020-ru}: with a subwavelength spacing, an array of the silver wires produces a homogeneous transverse magnetic field, $\vec{B}$, as shown in Fig.\,\ref{fig:fig1}b. Local excitation and probing of NVs within a pixel are possible by running a current through an individual wire.

\subsection{Metasurface design}
Here we discuss photonic design criteria to maximize the IR signal of spin ensemble sensors. The IR absorption readout has only been successfully implemented with bulk diamond samples due to the intrinsic absorption cross sectional area that is about an order of magnitude smaller than that of the triplet state transition \cite{Dumeige2013-wf,Jensen2014-xu}. This weak light-matter interaction can be enhanced by modifying the electromagnetic environment of quantum emitters. The rate of absorption of a quantum emitter under an oscillating electromagnetic field with frequency, $\omega_0$, can be expressed following Fermi's golden rule.
\begin{equation}\label{eq:eq1}
    \Gamma_\text{abs}=\frac{2\pi}{\hbar}|\braket{5|\vec{\mu}\cdot\vec{E}|6}|^2\rho(\omega_0)
\end{equation}
where $\vec{\mu}=e\cdot \vec{r}$ is the transition dipole moment operator, $\vec{E}$ is the electric fields, and $\rho(\omega)=\frac{1}{\pi\hbar}\frac{\frac{1}{2}\gamma^*}{(\omega-\omega_0)^2+(\frac{1}{2}\gamma^*)^2}$ is the electronic density of states, which is modeled as a continuum of final states with a Lorentzian distribution centered at $\omega_0$ with linewidth $\gamma^*$. The equation shows that the rate can be enhanced by increasing the electric field at the emitter position. Plasmonic structures can focus light intensity at subwavelength scales, and thus, they have  been used to increase spontaneous emission rates of single emitters or ensembles of emitters confined in a nanometer-scale volume \cite{Choy2013-bx,Akselrod2014-gx,Hoang2015-en,Karamlou2018-cv,Bogdanov2017-uu,De_Leon2012-ee,Hausmann2013-ob,Jun2010-wb,Russell2012-cn}. However, this field concentration comes with the trade-off of reducing the number of NV centers, $N_\text{NV}$, that are coupled to the optical field. Balancing this trade-off depends on the use case: sensing applications with a spatial resolution below the grating period benefit from highly localized SPP-like modes, while applications with larger lateral resolution benefit from more RWA-like modes that average over a larger $N_\text{NV}$. To guide the PQSM optimization, we adopt a figure of merit addressing the latter of $\braket{|E/E_0|^2}V_\text{pixel}n_\text{NV}$, where $\braket{|E/E_0|^2}=\int_\text{pixel}|E/E_0|^2dV/\int_\text{pixel}dV$ is the spatially averaged optical field enhancement over the single-pass field without plasmonic structure, $E_0$, $d_\text{NV}$ is the thickness of the diamond sensing layer with NV density of $n_\text{NV}$ , and $V_\text{pixel} =L^2 d_\text{NV}$  is the sensing volume.\par
The SPP-RWA resonance delocalizes the plasmonic modes, creating a large field enhancement within a few-micron-thick surface layer of diamond, as shown in the corresponding spatially-resolved electric field intensity profile (Fig.\,\ref{fig:fig1}b). The structure dimensions can be chosen to find the desired balance between SPP modes and RWA modes: stronger SPP localization resulting in more localized sensing can be traded against better sensitivity, and vice versa. SPPs do not couple with free-space light without satisfying the momentum matching conditions. An incoming far-field optical excitation can excite SPPs modes via a grating structure with period $p$ given by $G = 2 m/p$ where $m$ is an integer. To form a metallodielectric grating, the plasmonic nanostructures are arranged periodically with a period of $\lambda_0/n$, where $\lambda_0$ is 1042 nm and $n$ is the refractive index of diamond. As evident in the dispersion relation of RWA (Eq.\,\ref{eq:eq2}), this period satisfies the condition for first-diffraction-order RWA mode under normal incidence (i.e., $k_x = 0$). 
\begin{equation}\label{eq:eq2}
    \frac{\omega}{c}n=k_x+m\frac{2\pi}{p}
\end{equation}
where $c$ is the speed of light in vacuum, $k_x =k_0\sin(\theta_i)$ is the momentum component of free-space light in the direction of grating period, and $m$ denotes the diffraction order \cite{Gao2009-fk,Steele2003-nj}. When the RWA mode is excited, the incident electromagnetic wave diffracts parallel to the grating surface and creates a field profile that extends vertically away from the grating surface \cite{Wood1935-wq,Rayleigh1907-tp}. Thus, the SPP-RWA hybrid mode shows the electric field intensity profile extending a few microns from the grating surface while maintaining a large field concentration near the Ag-diamond interface as shown in  Fig.\,\ref{fig:fig1}b. While the RWA mode alone does not depend on the properties of the plasmonic material (Eq.\,\ref{eq:eq2}), the SPP-RWA hybrid mode is highly dependent on the dispersive permittivity of the plasmonic material. When silver is replaced with palladium (Pd), a weak plasmonic material, the electric field enhancement is heavily suppressed (refer to SI Section 1). As the SPP-RWA mode delocalizes the field away from the lossy material, this PQSM exhibits a quality factor of 935 at 1042 nm, an exceptionally high value for a plasmonic coupled mode. \par
To determine the signal from a pixel of the PQSM containing an ensemble of emitters, the rate of absorption is averaged over all four orientations of NV emitters. For a given angle, $\theta$, between the emitter’s transition dipole orientation and the electric field created by the PQSM in the diamond layer, Eq.\,\ref{eq:eq1} can be expressed as Eq.\,\ref{eq:eq3} in terms of the spontaneous emission rate of the singlet state transition, $\gamma=\frac{\omega_0^3|\braket{5|\vec{\mu}|6}|^2}{3\pi\epsilon\hbar c^3}$.
\begin{equation}\label{eq:eq3}
    \Gamma_\text{abs}=\frac{3}{\pi^2\hbar}\frac{\gamma}{\gamma^*}(\frac{\lambda_0}{n})^3\cdot\frac{1}{2}\epsilon_0\epsilon|\vec{E}|^2\cos^2(\theta)
\end{equation}
where $\epsilon$ is the relative permittivity of diamond. For a [100] diamond plane, all four orientations of NVs are expected to have equal contributions for the given SPP-RWA-induced field profile. The signal-to-noise ratio (SNR) of the pixelated plasmonic imaging surface is given by Eq.\,\ref{eq:eq4} under the assumption of the shot noise limit. 
\begin{equation}\label{eq:eq4}
\begin{split}
    \text{SNR}&=\frac{|N_0-N_1|}{\sqrt{N_0+N_1}} \\
    &=\sqrt{\frac{\Delta t_\text{mea}L^2}{\hbar\omega_0}}\frac{I_\text{out}(I_t,0)-I_\text{out}(I_t,\Omega_R)}{\sqrt{I_\text{out}(I_t,0)+I_\text{out}(I_t,\Omega_R)}}
\end{split}
\end{equation}
where $N_0$ and $N_1$ are the average numbers of photons detected from the m$_\text{s}$ = 0 and m$_\text{s}$ = $\pm1$ states, respectively, per measurement, $\Delta t_\text{mea}$ is the total readout time, and $I_\text{out}(I_t,\Omega_R)$ is the reflected intensity under green laser intensity, $I_t$, and  an applied microwave Rabi field, $\Omega_R$. In the limit of low contrast, the SNR can be re-written as follows:
\begin{equation}\label{eq:eq5}
\begin{split}
    \text{SNR}&=\sqrt{\frac{I_\text{out}(0,0)\Delta t_\text{mea}L^2}{2\hbar\omega_0}}(I_\text{NV}(I_t,\Omega_R)-I_\text{NV}(I_t,0)) \\
    &\propto \braket{|E/E_0|^2}V_\text{pixel}n_\text{NV}
\end{split}
\end{equation}
where $I_\text{NV}(I_t,\Omega_R)=\frac{I_\text{out}(0,0)-I_\text{out}(I_t,\Omega_R)}{I_\text{out}(0,0)}$ is the fractional change in IR intensity due to NV absorption, and $I_\text{NV}(I_t,\Omega_R)-I_\text{NV}(I_t,0)\propto\braket{|E/E_0|^2}V_\text{pixel}n_\text{NV}$ as described in detail in SI Section 3. Thus, the SNR scales with both spatially averaged electric field intensity enhancement factor and sensing volume  for a given NV density. The PQSM  combines a large field enhancement of the SPP mode and delocalization of the RWA mode, making the SPP-RWA hybrid mode well suited for ensemble-based sensing. 

\begin{figure*}
    \centering
    \includegraphics[width=0.8\textwidth]{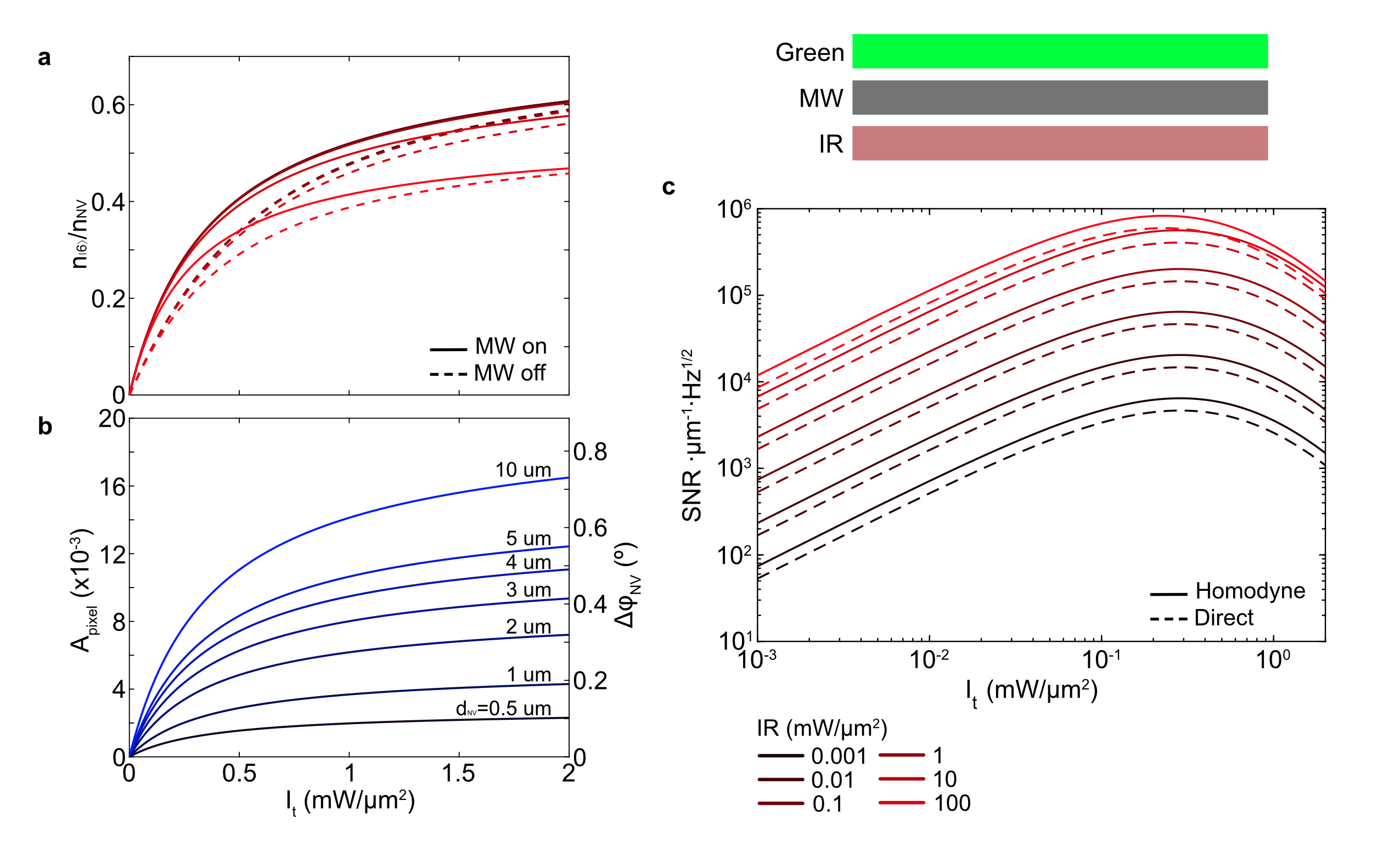}
    \caption{(a) The population of the ground singlet state as a function of 532 nm laser intensity with (solid) and without (dotted) an applied microwave field at a given sensing depth of 5 \textmu m. (b) Input-intensity-normalized NV absorption, $A_\text{pixel}$, per pixel (left y-axis) and corresponding phase changes of the spin-dependent reflected light, $\Delta\phi_\text{NV}$ (right y-axis). (c) SNR/$\sqrt{\Delta t_\text{mea}L^2}$ under steady state operation as a function of $I_t$ with varying $I_s$ for homodyne (solid) and direct (dotted) detection.}
    \label{fig:fig2}
\end{figure*}

\subsection{Metasurface spin-dependent response}

The spin-dependent IR absorption depends on the population of the ground singlet state, and it can be obtained by calculating the local density of each sublevel, $n_{\ket{i}}$, based on the coupled rate equations described in SI Section 2. 
As shown in Fig.\,\ref{fig:fig1}c, we use an eight-level rate equation model that accounts for photoionization. 
Under continuous wave (CW)-ODMR, Fig.\,\ref{fig:fig2}a shows the calculated $n_{\ket{6}}$, indicating that the population of the singlet states weakly depends on the IR probe intensity, $I_s$, until the absorption rate becomes comparable to the excited state decay rate. 
The SPP-RWA-induced field enhancement modifies the absorption rate as well as the radiative decay rate by $\gamma_\text{rad}\rightarrow F_p\gamma_\text{rad}+\gamma_\text{quenching}$, where $F_p$ is the Purcell factor. 
However, due to the low intrinsic quantum efficiency of the singlet state transition ($\frac{\gamma_\text{rad}}{\Gamma}\approx$0.1\% \cite{Ulbricht2018-yv}), the plasmonic structures have minimal effect on the overall excited state decay rate, $\Gamma$.  
Because the lifetime of the ground state is approximately two orders of magnitude longer than that of the excited state \cite{Acosta2010-cp}, the singlet state transition has an exceptionally high saturation intensity, enabling each NV  to absorb multiple photons per cycle. 
With resonant structures, the system can be brought to its saturation level at a lower incident intensity due to the transition rate enhanced by a factor of $\sim\braket{|E/E_0|^2}$. 
Similarly, the current PQSM structure can induce a grating mode resonant at 532 nm with an off-normal back illumination (refer to SI Section 1).

Under an applied microwave field, an increase in the population of the m$_\text{s} = \pm1$ states incurs an increase in absorption at 1042 nm (i.e.,
$I_\text{NV}(I_t,\Omega_R) > I_\text{NV}(I_t,0)$). 
This state-selective NV infrared absorption produces spin-dependent amplitude and phase changes in the IR probe field reflected from the PQSM. 
The input-intensity-normalized NV absorption, $A_\text{pixel}$, (equivalently, $I_\text{NV}(I_t,\Omega_R)\times\frac{I_\text{out}(0,0)}{I_s}$) with varying sensing depth is shown in Fig.\,\ref{fig:fig2}b. 
To account for stimulated emission, a net population (i.e., $n_{\ket{6}}$-$n_{\ket{5}}$) is used to calculate the net spin-dependent NV absorption. 
Based on the calculated $A_\text{pixel}$, the NV-induced corresponding phase change, $\Delta\phi_\text{NV}$, is obtained from the complex reflection coefficients simulated with FDTD method (Fig.\,\ref{fig:fig2}b).

The spin-dependent phase and amplitude changes of the signal allow for a phase-sensitive measurement.  
Here, we implement a phase-sensitive coherent homodyne detection, where a local oscillator interferes with the spin-dependent signal from the PQSM. A combination of $R$ and $\Delta\phi_\text{LO}$ is chosen to maximize the SNR (refer to SI Section 4) and the readout fidelity. The incident-intensity-normalized interfered intensity detected by the camera is given by Eq.\,\ref{eq:eq6}.
\begin{equation}\label{eq:eq6}
\begin{split}
    &\frac{I_\text{out}(I_t,\Omega_R,R,\Delta\phi_\text{LO})}{I_s}=(1-R)+R|r(I_t,\Omega_R)|^2 \\
    &+2\sqrt{(1-R)R}|r(I_t,\Omega_R)|\cos(\Delta\phi_\text{LO}+\Delta\phi_\text{NV}(I_t,\Omega_R))
\end{split}
\end{equation}
where $R$ is the power splitting ratio of the beam splitter, $r(I_t,\Omega_R)$ is the complex reflection coefficient of the PQSM, $\Delta\phi_\text{LO}$ is the relative phase difference between the LO and the reflected light of the PQSM when $I_t$ = 0 and $\Omega_R$ = 0, and $\Delta\phi_\text{NV}(I_t,\Omega_R)$ is an additional phase change incurred by the NV absorption.
Figure\,\ref{fig:fig2}c shows SNR that is normalized by the measurement time and pixel area of $L^2$. 
Under the conditions considered in this work, the photon shot noise dominates, and a better SNR is achieved by biasing the interferometric readout with a controlled phase difference. Homodyne detection is particularly advantageous for fast imaging on focal plane arrays. 

\section{Discussion}
\subsection{DC sensitivity}
The shot-noise-limited sensitivity of a CW-ODMR-based magnetometer per root area based on IR absorption measurement is given by Eq.\,\ref{eq:eq7}.\par
\begin{equation}\label{eq:eq7}
    \eta_\text{CW}^{A}=\frac{\hbar\Gamma_\text{MW}}{g\mu_\text{B}}\frac{\sqrt{\Delta t_\text{mea}L^2}}{\text{SNR}}
\end{equation}
where $g\approx2.003$ is the electronic g-factor of the NV center, $\mu_\text{B}$ is the Bohr magneton, and $\Gamma_\text{MW}$ is the magnetic-resonance linewidth which can be approximated as $\Gamma_\text{MW}=2/T_2^*$, assuming no power broadening from pump or microwaves. For a given NV density of $\sim$16 ppm, the dephasing time is limited by paramagnetic impurities with the conversion efficiency conservatively approximately as 16\% \cite{Ulbricht2018-yv}. An alternative magnetometry method to CW-ODMR, such as pulsed ODMR or Ramsey sequences, can be exploited to achieve $T_2^*$-limited performance.\par
\begin{figure}
    \centering
    \includegraphics[width=0.85\columnwidth]{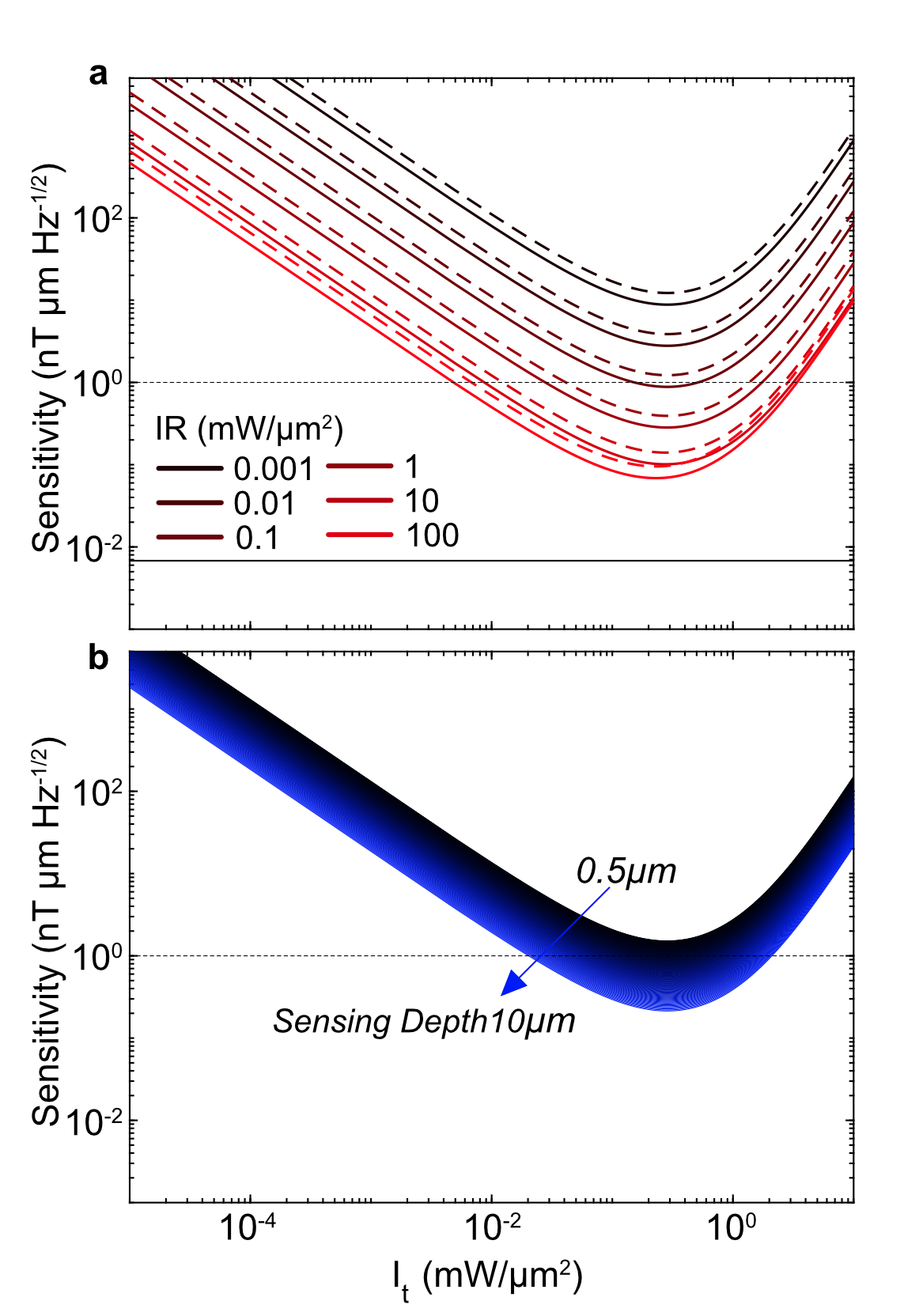}
    \caption{(a) Expected sensitivity per root sensing surface area with homodyne (solid) and direct  (dotted) detection as a function of $I_t$ with varying $I_s$ for a given $d_\text{NV}$ = 5 \textmu m. The dotted black line indicates 1 nT Hz$^{-\frac{1}{2}}$ sensitivity for a 1 \textmu m$^2$ sensing area. The solid black line indicates the spin-projection-noise-limited sensitivity. (b) Sensing-depth-dependent sensitivity with homodyne detection for a given $I_s$ = 1 mW/\textmu m$^2$.}
    \label{fig:fig3}
\end{figure}

It is useful to compare the photon-shot-noise-limited sensitivity with the spin-projection-noise-limited sensitivity of an ensemble magnetometer consisting of non-interacting spins.
\begin{equation}\label{eq:eq8}
    \eta_\text{sp}^{A,\text{ensemble}}=\frac{\hbar}{g\mu_\text{B}\sqrt{n_\text{NV}d_\text{NV}\tau}}
\end{equation}
where $\tau$ is the free precession time per measurement. Figure\,\ref{fig:fig3}a shows that the PQSM can achieve sub-nT Hz$^{-\frac{1}{2}}$ sensitivity per \textmu m$^2$ sensing surface area for a given an NV layer thickness of $d_\text{NV}$ = 5 \textmu m and remaining experimental  parameters listed in Table\,\ref{tab:table1}. As expected, sensitivity improves with increasing green laser intensity until two-photon-mediated  photo-ionization processes start to become considerable.  Furthermore, as shown in Fig.\,\ref{fig:fig3}b, increasing the sensing depth from 500 nm to 10 \textmu m improves the sensitivity by a factor of up to $\sim$ 9. There exists a trade-off between achievable sensitivity and spatial resolution.

\begin{table}\label{tab:table1}
\caption{Physical parameters used in this work.}
\begin{ruledtabular}
\begin{tabular} {lcc}
\textrm{Parameter}&
\textrm{Value}&
\textrm{Reference}\\
\colrule
$k_{31}$=$k_{42}$ & 66 \textmu s$^{-1}$ & \cite{Tetienne2012-ic} \\
$k_{35}$ & 7.9 \textmu s$^{-1}$ & \cite{Tetienne2012-ic} \\
$k_{45}$ & 53 \textmu s$^{-1}$ & \cite{Tetienne2012-ic} \\
$k_{61}$ & 1 \textmu s$^{-1}$ & \cite{Tetienne2012-ic} \\
$k_{62}$ & 0.7 \textmu s$^{-1}$ & \cite{Tetienne2012-ic} \\
$k_{38}$=$k_{48}$ & 41.8 MHz/mW & \cite{Tetienne2012-ic,Robledo2011-xg} \\
$k_{71}$=$k_{72}$ & 35.5 MHz/mW & \cite{Tetienne2012-ic,Robledo2011-xg} \\
$\Gamma$ & 1 ns$^{-1}$ & \cite{Acosta2010-cp}\\
$\Gamma_{\text{NV}^0}$ & 53 \textmu s$^{-1}$ & \cite{Tetienne2012-ic,Robledo2011-xg} \\
$\sigma_\text{t}$ & 3$\times$10$^{-21}$ m$^2$ & \cite{Wee2007-jo} \\
$\sigma_\text{s}$ & 3$\times$10$^{-22}$ m$^2$ & \cite{Dumeige2013-wf,Jensen2014-xu} \\
$\sigma_{\text{NV}^0}$ & 6$\times$10$^{-21}$ m$^2$ & \cite{Tetienne2012-ic,Robledo2011-xg} \\
$n_\text{NV}$ & 28$\times$10$^{23}$ m$^{-3}$ & \cite{Acosta2009-te} \\
$T_2^*$ & 200 ns & \cite{Acosta2009-te,Barry2020-ec} \\
$T_2$ & 2 \textmu s & \cite{Barry2020-ec} \\
$\Omega_R$ & 2$\pi$ $\times$ 1.5 MHz & \cite{Acosta2010-rp} \\
\end{tabular}
\end{ruledtabular}
\end{table}

\subsection{AC sensitivity}
\begin{figure*}
    \centering
    \includegraphics[width=0.8\textwidth]{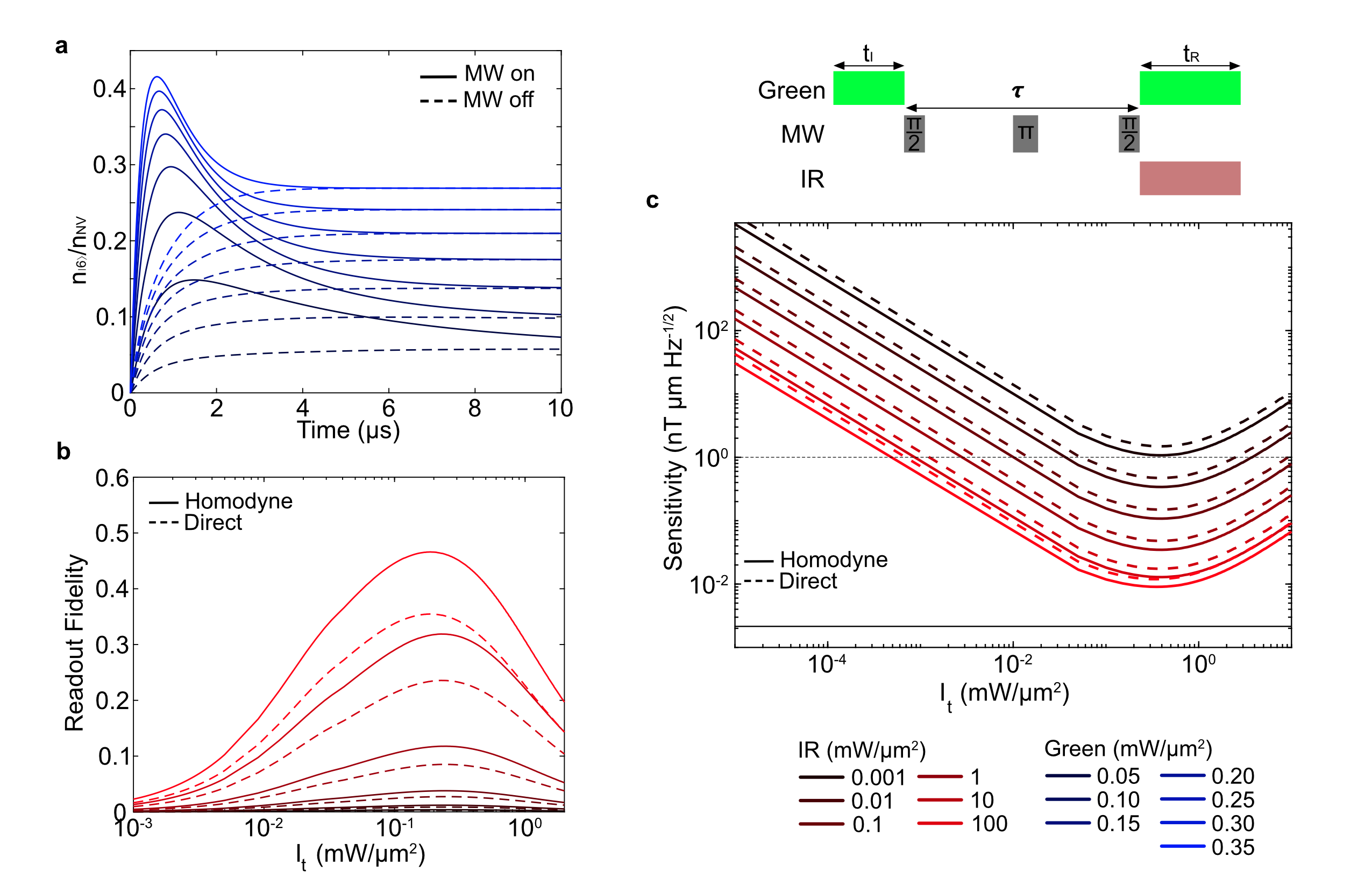}
    \caption{(a) Population of the singlet ground state with (solid) and without (dotted) an applied microwave field as a function of readout time at $d_\text{NV}$ = 5 \textmu m for a given $I_s$ = 1 mW/\textmu m$^2$. (b) Readout fidelity and (c) AC sensitivity per root sensing surface area as a function of $I_t$ with varying $I_s$ at $d_\text{NV}$ = 5 \textmu m for homodyne (solid) and direct (dotted) detection. The solid black line indicates the spin-projection-noise-limited sensitivity given by Eq.\,\ref{eq:eq8}.}
    \label{fig:fig4}
\end{figure*}

Sources of the NV spin dephasing can be largely eliminated with coherent control techniques such as the Hahn echo sequence. With an added $\pi$-pulse halfway through the interrogation time, a net phase accumulated due to a static or slowly varying magnetic field cancels out, and the interrogation time can be extended to a value of $\sim$ $T_2$. Thus, the AC sensitivity can improve by a factor of approximately $\sqrt{T_2^*/T_2}$ at the cost of a reduced bandwidth and insensitivity to magnetic field with an oscillating period longer than $T_2$. The sensitivity per root area for an ensemble-based AC magnetometer is given by Eq.\,\ref{eq:eq9} \cite{Barry2020-ec}.
\begin{equation}\label{eq:eq9}
    \eta_\text{a.c.}^{A}=\frac{\hbar\sigma_\text{R} e^{\tau/T_2}}{g\mu_\text{B}\sqrt{n_\text{NV}d_\text{NV}\tau}}\sqrt{1+\frac{t_\text{I}+t_\text{R}}{\tau}}
\end{equation}
where $T_2$ is the characteristic dephasing time, $t_\text{I}$ is the initialization time, $t_\text{R}$ is the readout time, and $\sigma_\text{R}$ is the readout fidelity. For a given NV density of $\sim$ 16 ppm, $T_2$ is about an order of magnitude longer than $T_2^*$ \cite{Barry2020-ec}. Rate equations are solved as a function of time to obtain time-dependent population evolution as shown in Fig.\,\ref{fig:fig4}a. The system loses spin polarization after a few microseconds of readout time. For given $I_t$ and $I_s$, an optimal readout time that maximizes the time-integrated signal is calculated and is shown to be near 500 ns (refer to SI Section 5). An additional shot noise introduced by the optical readout is quantified with the parameter $\sigma_\text{R}$ (Eq.\,\ref{eq:eq10}), which is equivalent to an inverse of readout fidelity \cite{Barry2020-ec}.
\begin{equation}\label{eq:eq10}
    \sigma_\text{R}=\sqrt{1+\frac{2(a+b)}{(a-b)^2}}
\end{equation}
where $a$ and $b$ are the average numbers of photons detected from the m$_\text{s}$ = 0 and m$_\text{s}$ = $\pm1$ states per spin per measurement, respectively. The PQSM achieves spin readout fidelity near 0.5 per shot. Figure\,\ref{fig:fig4}c shows the AC sensitivity down to 10 pT Hz$^{-\frac{1}{2}}$ per 1-\textmu m$^2$ sensing surface area. 

\section{Conclusion}
In summary, we report a diamond quantum sensing surface consisting of plasmonic nanostructures, which shows a sub-nT Hz$^{-\frac{1}{2}}$ sensitivity per a 1-\textmu m$^2$ sensing surface. This exceptional performance is achieved by the SPP-RWA resonance that optimizes an electric field enhancement within a micron-scale NV layer. The plasmonic structures of the PQSM also provide an optimal microwave control by generating a homogeneous magnetic field across a large sensing area. Combined with a homodyne detection, the PQSM makes a new type of quantum microscope that enables high-speed imaging measurements at the photon shot noise limit.\par
This PQSM has far-reaching implications in quantum science. The metasurface-coupled quantum emitter arrays can enable manipulation of accumulated phase and polarization at each position of a quantum emitter. By superposing spin-dependent reflection or transmission, it may even be possible to entangle different regions of the metasurface. The entangled quantum metasurface is useful in applications that demand measurements of correlated quantum fluctuation such as quantum spin liquids in quantum materials \cite{Meng2010-qp,Takagi2019-hl}. Such an approach can exploit entanglement-enhanced quantum sensing protocols to achieve performance beyond the standard quantum limit \cite{Cappellaro2009-os,Choi2017-lu}. 


\section{Acknowledgements}
L.K. acknowledges support through an appointment to the Intelligence Community Postdoctoral Research Fellowship Program at the Massachusetts Institute of Technology, administered by Oak Ridge Institute for Science and Education through an interagency agreement between the U.S. Department of Energy and the Office of the Director of National Intelligence. H. C. acknowledges support through Claude E. Shannon Fellowship and the DARPA DRINQS, D18AC00014 program. M.E.T. acknowledges support through the Army  Research Laboratory ENIAC Distinguished Postdoctoral Fellowship. D.E. acknowledges support from the Bose Research Fellowship, the Army Research Office W911NF-17-1-0435, and the NSF CUA. We thank Dr. Jennifer Schloss and Jordan Goldstein for helpful discussions.

\bibliography{references.bib}

\begin{thebibliography}{37}%
\makeatletter
\providecommand \@ifxundefined [1]{%
 \@ifx{#1\undefined}
}%
\providecommand \@ifnum [1]{%
 \ifnum #1\expandafter \@firstoftwo
 \else \expandafter \@secondoftwo
 \fi
}%
\providecommand \@ifx [1]{%
 \ifx #1\expandafter \@firstoftwo
 \else \expandafter \@secondoftwo
 \fi
}%
\providecommand \natexlab [1]{#1}%
\providecommand \enquote  [1]{``#1''}%
\providecommand \bibnamefont  [1]{#1}%
\providecommand \bibfnamefont [1]{#1}%
\providecommand \citenamefont [1]{#1}%
\providecommand \href@noop [0]{\@secondoftwo}%
\providecommand \href [0]{\begingroup \@sanitize@url \@href}%
\providecommand \@href[1]{\@@startlink{#1}\@@href}%
\providecommand \@@href[1]{\endgroup#1\@@endlink}%
\providecommand \@sanitize@url [0]{\catcode `\\12\catcode `\$12\catcode
  `\&12\catcode `\#12\catcode `\^12\catcode `\_12\catcode `\%12\relax}%
\providecommand \@@startlink[1]{}%
\providecommand \@@endlink[0]{}%
\providecommand \url  [0]{\begingroup\@sanitize@url \@url }%
\providecommand \@url [1]{\endgroup\@href {#1}{\urlprefix }}%
\providecommand \urlprefix  [0]{URL }%
\providecommand \Eprint [0]{\href }%
\providecommand \doibase [0]{http://dx.doi.org/}%
\providecommand \selectlanguage [0]{\@gobble}%
\providecommand \bibinfo  [0]{\@secondoftwo}%
\providecommand \bibfield  [0]{\@secondoftwo}%
\providecommand \translation [1]{[#1]}%
\providecommand \BibitemOpen [0]{}%
\providecommand \bibitemStop [0]{}%
\providecommand \bibitemNoStop [0]{.\EOS\space}%
\providecommand \EOS [0]{\spacefactor3000\relax}%
\providecommand \BibitemShut  [1]{\csname bibitem#1\endcsname}%
\let\auto@bib@innerbib\@empty
\bibitem [{\citenamefont {Acosta}\ \emph
  {et~al.}(2010{\natexlab{a}})\citenamefont {Acosta}, \citenamefont {Bauch},
  \citenamefont {Jarmola}, \citenamefont {Zipp}, \citenamefont {Ledbetter},\
  and\ \citenamefont {Budker}}]{Acosta2010-rp}%
  \BibitemOpen
  \bibfield  {author} {\bibinfo {author} {\bibfnamefont {V.~M.}\ \bibnamefont
  {Acosta}}, \bibinfo {author} {\bibfnamefont {E.}~\bibnamefont {Bauch}},
  \bibinfo {author} {\bibfnamefont {A.}~\bibnamefont {Jarmola}}, \bibinfo
  {author} {\bibfnamefont {L.~J.}\ \bibnamefont {Zipp}}, \bibinfo {author}
  {\bibfnamefont {M.~P.}\ \bibnamefont {Ledbetter}}, \ and\ \bibinfo {author}
  {\bibfnamefont {D.}~\bibnamefont {Budker}},\ }\href@noop {} {\bibfield
  {journal} {\bibinfo  {journal} {Appl. Phys. Lett.}\ }\textbf {\bibinfo
  {volume} {97}} (\bibinfo {year} {2010}{\natexlab{a}})}\BibitemShut {NoStop}%
\bibitem [{\citenamefont {Bougas}\ \emph {et~al.}(2018)\citenamefont {Bougas},
  \citenamefont {Wilzewski}, \citenamefont {Dumeige}, \citenamefont {Antypas},
  \citenamefont {Wu}, \citenamefont {Wickenbrock}, \citenamefont {Bourgeois},
  \citenamefont {Nesladek}, \citenamefont {Clevenson}, \citenamefont {Braje},
  \citenamefont {Englund},\ and\ \citenamefont {Budker}}]{Bougas2018-ej}%
  \BibitemOpen
  \bibfield  {author} {\bibinfo {author} {\bibfnamefont {L.}~\bibnamefont
  {Bougas}}, \bibinfo {author} {\bibfnamefont {A.}~\bibnamefont {Wilzewski}},
  \bibinfo {author} {\bibfnamefont {Y.}~\bibnamefont {Dumeige}}, \bibinfo
  {author} {\bibfnamefont {D.}~\bibnamefont {Antypas}}, \bibinfo {author}
  {\bibfnamefont {T.}~\bibnamefont {Wu}}, \bibinfo {author} {\bibfnamefont
  {A.}~\bibnamefont {Wickenbrock}}, \bibinfo {author} {\bibfnamefont
  {E.}~\bibnamefont {Bourgeois}}, \bibinfo {author} {\bibfnamefont
  {M.}~\bibnamefont {Nesladek}}, \bibinfo {author} {\bibfnamefont
  {H.}~\bibnamefont {Clevenson}}, \bibinfo {author} {\bibfnamefont
  {D.}~\bibnamefont {Braje}}, \bibinfo {author} {\bibfnamefont
  {D.}~\bibnamefont {Englund}}, \ and\ \bibinfo {author} {\bibfnamefont
  {D.}~\bibnamefont {Budker}},\ }\href@noop {} {\bibfield  {journal} {\bibinfo
  {journal} {Micromachines (Basel)}\ }\textbf {\bibinfo {volume} {9}} (\bibinfo
  {year} {2018})}\BibitemShut {NoStop}%
\bibitem [{\citenamefont {Dumeige}\ \emph {et~al.}(2013)\citenamefont
  {Dumeige}, \citenamefont {Chipaux}, \citenamefont {Jacques}, \citenamefont
  {Treussart}, \citenamefont {Roch}, \citenamefont {Debuisschert},
  \citenamefont {Acosta}, \citenamefont {Jarmola}, \citenamefont {Jensen},
  \citenamefont {Kehayias},\ and\ \citenamefont {Budker}}]{Dumeige2013-wf}%
  \BibitemOpen
  \bibfield  {author} {\bibinfo {author} {\bibfnamefont {Y.}~\bibnamefont
  {Dumeige}}, \bibinfo {author} {\bibfnamefont {M.}~\bibnamefont {Chipaux}},
  \bibinfo {author} {\bibfnamefont {V.}~\bibnamefont {Jacques}}, \bibinfo
  {author} {\bibfnamefont {F.}~\bibnamefont {Treussart}}, \bibinfo {author}
  {\bibfnamefont {J.-F.}\ \bibnamefont {Roch}}, \bibinfo {author}
  {\bibfnamefont {T.}~\bibnamefont {Debuisschert}}, \bibinfo {author}
  {\bibfnamefont {V.~M.}\ \bibnamefont {Acosta}}, \bibinfo {author}
  {\bibfnamefont {A.}~\bibnamefont {Jarmola}}, \bibinfo {author} {\bibfnamefont
  {K.}~\bibnamefont {Jensen}}, \bibinfo {author} {\bibfnamefont
  {P.}~\bibnamefont {Kehayias}}, \ and\ \bibinfo {author} {\bibfnamefont
  {D.}~\bibnamefont {Budker}},\ }\href@noop {} {\bibfield  {journal} {\bibinfo
  {journal} {Phys. Rev. B Condens. Matter}\ }\textbf {\bibinfo {volume} {87}},\
  \bibinfo {pages} {155202} (\bibinfo {year} {2013})}\BibitemShut {NoStop}%
\bibitem [{\citenamefont {Barry}\ \emph {et~al.}(2016)\citenamefont {Barry},
  \citenamefont {Turner}, \citenamefont {Schloss}, \citenamefont {Glenn},
  \citenamefont {Song}, \citenamefont {Lukin}, \citenamefont {Park},\ and\
  \citenamefont {Walsworth}}]{Barry2016-rc}%
  \BibitemOpen
  \bibfield  {author} {\bibinfo {author} {\bibfnamefont {J.~F.}\ \bibnamefont
  {Barry}}, \bibinfo {author} {\bibfnamefont {M.~J.}\ \bibnamefont {Turner}},
  \bibinfo {author} {\bibfnamefont {J.~M.}\ \bibnamefont {Schloss}}, \bibinfo
  {author} {\bibfnamefont {D.~R.}\ \bibnamefont {Glenn}}, \bibinfo {author}
  {\bibfnamefont {Y.}~\bibnamefont {Song}}, \bibinfo {author} {\bibfnamefont
  {M.~D.}\ \bibnamefont {Lukin}}, \bibinfo {author} {\bibfnamefont
  {H.}~\bibnamefont {Park}}, \ and\ \bibinfo {author} {\bibfnamefont {R.~L.}\
  \bibnamefont {Walsworth}},\ }\href@noop {} {\bibfield  {journal} {\bibinfo
  {journal} {Proceedings of the National Academy of Sciences}\ }\textbf
  {\bibinfo {volume} {113}},\ \bibinfo {pages} {14133} (\bibinfo {year}
  {2016})}\BibitemShut {NoStop}%
\bibitem [{\citenamefont {Turner}\ \emph {et~al.}(2020)\citenamefont {Turner},
  \citenamefont {Langellier}, \citenamefont {Bainbridge}, \citenamefont
  {Walters}, \citenamefont {Meesala}, \citenamefont {Babinec}, \citenamefont
  {Kehayias}, \citenamefont {Yacoby}, \citenamefont {Hu}, \citenamefont {Lon{\v
  c}ar}, \citenamefont {Walsworth},\ and\ \citenamefont
  {Levine}}]{Turner2020-rq}%
  \BibitemOpen
  \bibfield  {author} {\bibinfo {author} {\bibfnamefont {M.~J.}\ \bibnamefont
  {Turner}}, \bibinfo {author} {\bibfnamefont {N.}~\bibnamefont {Langellier}},
  \bibinfo {author} {\bibfnamefont {R.}~\bibnamefont {Bainbridge}}, \bibinfo
  {author} {\bibfnamefont {D.}~\bibnamefont {Walters}}, \bibinfo {author}
  {\bibfnamefont {S.}~\bibnamefont {Meesala}}, \bibinfo {author} {\bibfnamefont
  {T.~M.}\ \bibnamefont {Babinec}}, \bibinfo {author} {\bibfnamefont
  {P.}~\bibnamefont {Kehayias}}, \bibinfo {author} {\bibfnamefont
  {A.}~\bibnamefont {Yacoby}}, \bibinfo {author} {\bibfnamefont
  {E.}~\bibnamefont {Hu}}, \bibinfo {author} {\bibfnamefont {M.}~\bibnamefont
  {Lon{\v c}ar}}, \bibinfo {author} {\bibfnamefont {R.~L.}\ \bibnamefont
  {Walsworth}}, \ and\ \bibinfo {author} {\bibfnamefont {E.~V.}\ \bibnamefont
  {Levine}},\ }\href@noop {} {\bibfield  {journal} {\bibinfo  {journal}
  {arXiv}\ } (\bibinfo {year} {2020})},\ \Eprint
  {http://arxiv.org/abs/2004.03707} {arXiv:2004.03707 [quant-ph]} \BibitemShut
  {NoStop}%
\bibitem [{\citenamefont {Thiel}\ \emph {et~al.}(2016)\citenamefont {Thiel},
  \citenamefont {Rohner}, \citenamefont {Ganzhorn}, \citenamefont {Appel},
  \citenamefont {Neu}, \citenamefont {M{\"u}ller}, \citenamefont {Kleiner},
  \citenamefont {Koelle},\ and\ \citenamefont {Maletinsky}}]{Thiel2016-yf}%
  \BibitemOpen
  \bibfield  {author} {\bibinfo {author} {\bibfnamefont {L.}~\bibnamefont
  {Thiel}}, \bibinfo {author} {\bibfnamefont {D.}~\bibnamefont {Rohner}},
  \bibinfo {author} {\bibfnamefont {M.}~\bibnamefont {Ganzhorn}}, \bibinfo
  {author} {\bibfnamefont {P.}~\bibnamefont {Appel}}, \bibinfo {author}
  {\bibfnamefont {E.}~\bibnamefont {Neu}}, \bibinfo {author} {\bibfnamefont
  {B.}~\bibnamefont {M{\"u}ller}}, \bibinfo {author} {\bibfnamefont
  {R.}~\bibnamefont {Kleiner}}, \bibinfo {author} {\bibfnamefont
  {D.}~\bibnamefont {Koelle}}, \ and\ \bibinfo {author} {\bibfnamefont
  {P.}~\bibnamefont {Maletinsky}},\ }\href@noop {} {\bibfield  {journal}
  {\bibinfo  {journal} {Nat. Nanotechnol.}\ }\textbf {\bibinfo {volume} {11}},\
  \bibinfo {pages} {677} (\bibinfo {year} {2016})}\BibitemShut {NoStop}%
\bibitem [{\citenamefont {Tetienne}\ \emph {et~al.}(2015)\citenamefont
  {Tetienne}, \citenamefont {Hingant}, \citenamefont {Mart{\'\i}nez},
  \citenamefont {Rohart}, \citenamefont {Thiaville}, \citenamefont {Diez},
  \citenamefont {Garcia}, \citenamefont {Adam}, \citenamefont {Kim},
  \citenamefont {Roch}, \citenamefont {Miron}, \citenamefont {Gaudin},
  \citenamefont {Vila}, \citenamefont {Ocker}, \citenamefont {Ravelosona},\
  and\ \citenamefont {Jacques}}]{Tetienne2015-uv}%
  \BibitemOpen
  \bibfield  {author} {\bibinfo {author} {\bibfnamefont {J.-P.}\ \bibnamefont
  {Tetienne}}, \bibinfo {author} {\bibfnamefont {T.}~\bibnamefont {Hingant}},
  \bibinfo {author} {\bibfnamefont {L.~J.}\ \bibnamefont {Mart{\'\i}nez}},
  \bibinfo {author} {\bibfnamefont {S.}~\bibnamefont {Rohart}}, \bibinfo
  {author} {\bibfnamefont {A.}~\bibnamefont {Thiaville}}, \bibinfo {author}
  {\bibfnamefont {L.~H.}\ \bibnamefont {Diez}}, \bibinfo {author}
  {\bibfnamefont {K.}~\bibnamefont {Garcia}}, \bibinfo {author} {\bibfnamefont
  {J.-P.}\ \bibnamefont {Adam}}, \bibinfo {author} {\bibfnamefont {J.-V.}\
  \bibnamefont {Kim}}, \bibinfo {author} {\bibfnamefont {J.-F.}\ \bibnamefont
  {Roch}}, \bibinfo {author} {\bibfnamefont {I.~M.}\ \bibnamefont {Miron}},
  \bibinfo {author} {\bibfnamefont {G.}~\bibnamefont {Gaudin}}, \bibinfo
  {author} {\bibfnamefont {L.}~\bibnamefont {Vila}}, \bibinfo {author}
  {\bibfnamefont {B.}~\bibnamefont {Ocker}}, \bibinfo {author} {\bibfnamefont
  {D.}~\bibnamefont {Ravelosona}}, \ and\ \bibinfo {author} {\bibfnamefont
  {V.}~\bibnamefont {Jacques}},\ }\href@noop {} {\bibfield  {journal} {\bibinfo
   {journal} {Nat. Commun.}\ }\textbf {\bibinfo {volume} {6}},\ \bibinfo
  {pages} {6733} (\bibinfo {year} {2015})}\BibitemShut {NoStop}%
\bibitem [{\citenamefont {Du}\ \emph {et~al.}(2017)\citenamefont {Du},
  \citenamefont {van~der Sar}, \citenamefont {Zhou}, \citenamefont {Upadhyaya},
  \citenamefont {Casola}, \citenamefont {Zhang}, \citenamefont {Onbasli},
  \citenamefont {Ross}, \citenamefont {Walsworth}, \citenamefont
  {Tserkovnyak},\ and\ \citenamefont {Yacoby}}]{Du2017-xr}%
  \BibitemOpen
  \bibfield  {author} {\bibinfo {author} {\bibfnamefont {C.}~\bibnamefont
  {Du}}, \bibinfo {author} {\bibfnamefont {T.}~\bibnamefont {van~der Sar}},
  \bibinfo {author} {\bibfnamefont {T.~X.}\ \bibnamefont {Zhou}}, \bibinfo
  {author} {\bibfnamefont {P.}~\bibnamefont {Upadhyaya}}, \bibinfo {author}
  {\bibfnamefont {F.}~\bibnamefont {Casola}}, \bibinfo {author} {\bibfnamefont
  {H.}~\bibnamefont {Zhang}}, \bibinfo {author} {\bibfnamefont {M.~C.}\
  \bibnamefont {Onbasli}}, \bibinfo {author} {\bibfnamefont {C.~A.}\
  \bibnamefont {Ross}}, \bibinfo {author} {\bibfnamefont {R.~L.}\ \bibnamefont
  {Walsworth}}, \bibinfo {author} {\bibfnamefont {Y.}~\bibnamefont
  {Tserkovnyak}}, \ and\ \bibinfo {author} {\bibfnamefont {A.}~\bibnamefont
  {Yacoby}},\ }\href@noop {} {\bibfield  {journal} {\bibinfo  {journal}
  {Science}\ }\textbf {\bibinfo {volume} {357}},\ \bibinfo {pages} {195}
  (\bibinfo {year} {2017})}\BibitemShut {NoStop}%
\bibitem [{\citenamefont {Ku}\ \emph {et~al.}(2020)\citenamefont {Ku},
  \citenamefont {Zhou}, \citenamefont {Li}, \citenamefont {Shin}, \citenamefont
  {Shi}, \citenamefont {Burch}, \citenamefont {Anderson}, \citenamefont
  {Pierce}, \citenamefont {Xie}, \citenamefont {Hamo}, \citenamefont {Vool},
  \citenamefont {Zhang}, \citenamefont {Casola}, \citenamefont {Taniguchi},
  \citenamefont {Watanabe}, \citenamefont {Fogler}, \citenamefont {Kim},
  \citenamefont {Yacoby},\ and\ \citenamefont {Walsworth}}]{Ku2020-ev}%
  \BibitemOpen
  \bibfield  {author} {\bibinfo {author} {\bibfnamefont {M.~J.~H.}\
  \bibnamefont {Ku}}, \bibinfo {author} {\bibfnamefont {T.~X.}\ \bibnamefont
  {Zhou}}, \bibinfo {author} {\bibfnamefont {Q.}~\bibnamefont {Li}}, \bibinfo
  {author} {\bibfnamefont {Y.~J.}\ \bibnamefont {Shin}}, \bibinfo {author}
  {\bibfnamefont {J.~K.}\ \bibnamefont {Shi}}, \bibinfo {author} {\bibfnamefont
  {C.}~\bibnamefont {Burch}}, \bibinfo {author} {\bibfnamefont {L.~E.}\
  \bibnamefont {Anderson}}, \bibinfo {author} {\bibfnamefont {A.~T.}\
  \bibnamefont {Pierce}}, \bibinfo {author} {\bibfnamefont {Y.}~\bibnamefont
  {Xie}}, \bibinfo {author} {\bibfnamefont {A.}~\bibnamefont {Hamo}}, \bibinfo
  {author} {\bibfnamefont {U.}~\bibnamefont {Vool}}, \bibinfo {author}
  {\bibfnamefont {H.}~\bibnamefont {Zhang}}, \bibinfo {author} {\bibfnamefont
  {F.}~\bibnamefont {Casola}}, \bibinfo {author} {\bibfnamefont
  {T.}~\bibnamefont {Taniguchi}}, \bibinfo {author} {\bibfnamefont
  {K.}~\bibnamefont {Watanabe}}, \bibinfo {author} {\bibfnamefont {M.~M.}\
  \bibnamefont {Fogler}}, \bibinfo {author} {\bibfnamefont {P.}~\bibnamefont
  {Kim}}, \bibinfo {author} {\bibfnamefont {A.}~\bibnamefont {Yacoby}}, \ and\
  \bibinfo {author} {\bibfnamefont {R.~L.}\ \bibnamefont {Walsworth}},\
  }\href@noop {} {\bibfield  {journal} {\bibinfo  {journal} {Nature}\ }\textbf
  {\bibinfo {volume} {583}},\ \bibinfo {pages} {537} (\bibinfo {year}
  {2020})}\BibitemShut {NoStop}%
\bibitem [{\citenamefont {Barry}\ \emph {et~al.}(2020)\citenamefont {Barry},
  \citenamefont {Schloss}, \citenamefont {Bauch}, \citenamefont {Turner},
  \citenamefont {Hart}, \citenamefont {Pham},\ and\ \citenamefont
  {Walsworth}}]{Barry2020-ec}%
  \BibitemOpen
  \bibfield  {author} {\bibinfo {author} {\bibfnamefont {J.~F.}\ \bibnamefont
  {Barry}}, \bibinfo {author} {\bibfnamefont {J.~M.}\ \bibnamefont {Schloss}},
  \bibinfo {author} {\bibfnamefont {E.}~\bibnamefont {Bauch}}, \bibinfo
  {author} {\bibfnamefont {M.~J.}\ \bibnamefont {Turner}}, \bibinfo {author}
  {\bibfnamefont {C.~A.}\ \bibnamefont {Hart}}, \bibinfo {author}
  {\bibfnamefont {L.~M.}\ \bibnamefont {Pham}}, \ and\ \bibinfo {author}
  {\bibfnamefont {R.~L.}\ \bibnamefont {Walsworth}},\ }\href@noop {} {\bibfield
   {journal} {\bibinfo  {journal} {Rev. Mod. Phys.}\ }\textbf {\bibinfo
  {volume} {92}},\ \bibinfo {pages} {015004} (\bibinfo {year}
  {2020})}\BibitemShut {NoStop}%
\bibitem [{\citenamefont {Chatzidrosos}\ \emph {et~al.}(2017)\citenamefont
  {Chatzidrosos}, \citenamefont {Wickenbrock}, \citenamefont {Bougas},
  \citenamefont {Leefer}, \citenamefont {Wu}, \citenamefont {Jensen},
  \citenamefont {Dumeige},\ and\ \citenamefont {Budker}}]{Chatzidrosos2017-vf}%
  \BibitemOpen
  \bibfield  {author} {\bibinfo {author} {\bibfnamefont {G.}~\bibnamefont
  {Chatzidrosos}}, \bibinfo {author} {\bibfnamefont {A.}~\bibnamefont
  {Wickenbrock}}, \bibinfo {author} {\bibfnamefont {L.}~\bibnamefont {Bougas}},
  \bibinfo {author} {\bibfnamefont {N.}~\bibnamefont {Leefer}}, \bibinfo
  {author} {\bibfnamefont {T.}~\bibnamefont {Wu}}, \bibinfo {author}
  {\bibfnamefont {K.}~\bibnamefont {Jensen}}, \bibinfo {author} {\bibfnamefont
  {Y.}~\bibnamefont {Dumeige}}, \ and\ \bibinfo {author} {\bibfnamefont
  {D.}~\bibnamefont {Budker}},\ }\href@noop {} {\bibfield  {journal} {\bibinfo
  {journal} {Phys. Rev. Applied}\ }\textbf {\bibinfo {volume} {8}},\ \bibinfo
  {pages} {044019} (\bibinfo {year} {2017})}\BibitemShut {NoStop}%
\bibitem [{\citenamefont {Jensen}\ \emph {et~al.}(2014)\citenamefont {Jensen},
  \citenamefont {Leefer}, \citenamefont {Jarmola}, \citenamefont {Dumeige},
  \citenamefont {Acosta}, \citenamefont {Kehayias}, \citenamefont {Patton},\
  and\ \citenamefont {Budker}}]{Jensen2014-xu}%
  \BibitemOpen
  \bibfield  {author} {\bibinfo {author} {\bibfnamefont {K.}~\bibnamefont
  {Jensen}}, \bibinfo {author} {\bibfnamefont {N.}~\bibnamefont {Leefer}},
  \bibinfo {author} {\bibfnamefont {A.}~\bibnamefont {Jarmola}}, \bibinfo
  {author} {\bibfnamefont {Y.}~\bibnamefont {Dumeige}}, \bibinfo {author}
  {\bibfnamefont {V.~M.}\ \bibnamefont {Acosta}}, \bibinfo {author}
  {\bibfnamefont {P.}~\bibnamefont {Kehayias}}, \bibinfo {author}
  {\bibfnamefont {B.}~\bibnamefont {Patton}}, \ and\ \bibinfo {author}
  {\bibfnamefont {D.}~\bibnamefont {Budker}},\ }\href@noop {} {\bibfield
  {journal} {\bibinfo  {journal} {Phys. Rev. Lett.}\ }\textbf {\bibinfo
  {volume} {112}},\ \bibinfo {pages} {160802} (\bibinfo {year}
  {2014})}\BibitemShut {NoStop}%
\bibitem [{\citenamefont {Gao}\ \emph {et~al.}(2009)\citenamefont {Gao},
  \citenamefont {McMahon}, \citenamefont {Lee}, \citenamefont {Henzie},
  \citenamefont {Gray}, \citenamefont {Schatz},\ and\ \citenamefont
  {Odom}}]{Gao2009-fk}%
  \BibitemOpen
  \bibfield  {author} {\bibinfo {author} {\bibfnamefont {H.}~\bibnamefont
  {Gao}}, \bibinfo {author} {\bibfnamefont {J.~M.}\ \bibnamefont {McMahon}},
  \bibinfo {author} {\bibfnamefont {M.~H.}\ \bibnamefont {Lee}}, \bibinfo
  {author} {\bibfnamefont {J.}~\bibnamefont {Henzie}}, \bibinfo {author}
  {\bibfnamefont {S.~K.}\ \bibnamefont {Gray}}, \bibinfo {author}
  {\bibfnamefont {G.~C.}\ \bibnamefont {Schatz}}, \ and\ \bibinfo {author}
  {\bibfnamefont {T.~W.}\ \bibnamefont {Odom}},\ }\href@noop {} {\bibfield
  {journal} {\bibinfo  {journal} {Opt. Express}\ }\textbf {\bibinfo {volume}
  {17}},\ \bibinfo {pages} {2334} (\bibinfo {year} {2009})}\BibitemShut
  {NoStop}%
\bibitem [{\citenamefont {Steele}\ \emph {et~al.}(2003)\citenamefont {Steele},
  \citenamefont {Moran}, \citenamefont {Lee}, \citenamefont {Aguirre},\ and\
  \citenamefont {Halas}}]{Steele2003-nj}%
  \BibitemOpen
  \bibfield  {author} {\bibinfo {author} {\bibfnamefont {J.~M.}\ \bibnamefont
  {Steele}}, \bibinfo {author} {\bibfnamefont {C.~E.}\ \bibnamefont {Moran}},
  \bibinfo {author} {\bibfnamefont {A.}~\bibnamefont {Lee}}, \bibinfo {author}
  {\bibfnamefont {C.~M.}\ \bibnamefont {Aguirre}}, \ and\ \bibinfo {author}
  {\bibfnamefont {N.~J.}\ \bibnamefont {Halas}},\ }\href@noop {} {\bibfield
  {journal} {\bibinfo  {journal} {Phys. Rev. B Condens. Matter}\ }\textbf
  {\bibinfo {volume} {68}},\ \bibinfo {pages} {205103} (\bibinfo {year}
  {2003})}\BibitemShut {NoStop}%
\bibitem [{\citenamefont {Acosta}\ \emph
  {et~al.}(2010{\natexlab{b}})\citenamefont {Acosta}, \citenamefont {Jarmola},
  \citenamefont {Bauch},\ and\ \citenamefont {Budker}}]{Acosta2010-cp}%
  \BibitemOpen
  \bibfield  {author} {\bibinfo {author} {\bibfnamefont {V.~M.}\ \bibnamefont
  {Acosta}}, \bibinfo {author} {\bibfnamefont {A.}~\bibnamefont {Jarmola}},
  \bibinfo {author} {\bibfnamefont {E.}~\bibnamefont {Bauch}}, \ and\ \bibinfo
  {author} {\bibfnamefont {D.}~\bibnamefont {Budker}},\ }\href@noop {}
  {\bibfield  {journal} {\bibinfo  {journal} {Phys. Rev. B Condens. Matter}\
  }\textbf {\bibinfo {volume} {82}},\ \bibinfo {pages} {201202} (\bibinfo
  {year} {2010}{\natexlab{b}})}\BibitemShut {NoStop}%
\bibitem [{\citenamefont {Shalaginov}\ \emph {et~al.}(2020)\citenamefont
  {Shalaginov}, \citenamefont {Bogdanov}, \citenamefont {Lagutchev},
  \citenamefont {Kildishev}, \citenamefont {Boltasseva},\ and\ \citenamefont
  {Shalaev}}]{Shalaginov2020-sd}%
  \BibitemOpen
  \bibfield  {author} {\bibinfo {author} {\bibfnamefont {M.~Y.}\ \bibnamefont
  {Shalaginov}}, \bibinfo {author} {\bibfnamefont {S.~I.}\ \bibnamefont
  {Bogdanov}}, \bibinfo {author} {\bibfnamefont {A.~S.}\ \bibnamefont
  {Lagutchev}}, \bibinfo {author} {\bibfnamefont {A.~V.}\ \bibnamefont
  {Kildishev}}, \bibinfo {author} {\bibfnamefont {A.}~\bibnamefont
  {Boltasseva}}, \ and\ \bibinfo {author} {\bibfnamefont {V.~M.}\ \bibnamefont
  {Shalaev}},\ }\href@noop {} {\bibfield  {journal} {\bibinfo  {journal} {ACS
  Photonics}\ }\textbf {\bibinfo {volume} {7}},\ \bibinfo {pages} {2018}
  (\bibinfo {year} {2020})}\BibitemShut {NoStop}%
\bibitem [{\citenamefont {Ibrahim}\ \emph {et~al.}(2020)\citenamefont
  {Ibrahim}, \citenamefont {Foy}, \citenamefont {Englund},\ and\ \citenamefont
  {Han}}]{Ibrahim2020-ru}%
  \BibitemOpen
  \bibfield  {author} {\bibinfo {author} {\bibfnamefont {M.~I.}\ \bibnamefont
  {Ibrahim}}, \bibinfo {author} {\bibfnamefont {C.}~\bibnamefont {Foy}},
  \bibinfo {author} {\bibfnamefont {D.~R.}\ \bibnamefont {Englund}}, \ and\
  \bibinfo {author} {\bibfnamefont {R.}~\bibnamefont {Han}},\ }\href@noop {} {\
   (\bibinfo {year} {2020})},\ \Eprint {http://arxiv.org/abs/2005.05638}
  {arXiv:2005.05638 [physics.app-ph]} \BibitemShut {NoStop}%
\bibitem [{\citenamefont {Choy}\ \emph {et~al.}(2013)\citenamefont {Choy},
  \citenamefont {Bulu}, \citenamefont {Hausmann}, \citenamefont {Janitz},
  \citenamefont {Huang},\ and\ \citenamefont {Lon{\v c}ar}}]{Choy2013-bx}%
  \BibitemOpen
  \bibfield  {author} {\bibinfo {author} {\bibfnamefont {J.~T.}\ \bibnamefont
  {Choy}}, \bibinfo {author} {\bibfnamefont {I.}~\bibnamefont {Bulu}}, \bibinfo
  {author} {\bibfnamefont {B.~J.~M.}\ \bibnamefont {Hausmann}}, \bibinfo
  {author} {\bibfnamefont {E.}~\bibnamefont {Janitz}}, \bibinfo {author}
  {\bibfnamefont {I.-C.}\ \bibnamefont {Huang}}, \ and\ \bibinfo {author}
  {\bibfnamefont {M.}~\bibnamefont {Lon{\v c}ar}},\ }\href@noop {} {\bibfield
  {journal} {\bibinfo  {journal} {Appl. Phys. Lett.}\ }\textbf {\bibinfo
  {volume} {103}},\ \bibinfo {pages} {161101} (\bibinfo {year}
  {2013})}\BibitemShut {NoStop}%
\bibitem [{\citenamefont {Akselrod}\ \emph {et~al.}(2014)\citenamefont
  {Akselrod}, \citenamefont {Argyropoulos}, \citenamefont {Hoang},
  \citenamefont {Cirac{\`\i}}, \citenamefont {Fang}, \citenamefont {Huang},
  \citenamefont {Smith},\ and\ \citenamefont {Mikkelsen}}]{Akselrod2014-gx}%
  \BibitemOpen
  \bibfield  {author} {\bibinfo {author} {\bibfnamefont {G.~M.}\ \bibnamefont
  {Akselrod}}, \bibinfo {author} {\bibfnamefont {C.}~\bibnamefont
  {Argyropoulos}}, \bibinfo {author} {\bibfnamefont {T.~B.}\ \bibnamefont
  {Hoang}}, \bibinfo {author} {\bibfnamefont {C.}~\bibnamefont {Cirac{\`\i}}},
  \bibinfo {author} {\bibfnamefont {C.}~\bibnamefont {Fang}}, \bibinfo {author}
  {\bibfnamefont {J.}~\bibnamefont {Huang}}, \bibinfo {author} {\bibfnamefont
  {D.~R.}\ \bibnamefont {Smith}}, \ and\ \bibinfo {author} {\bibfnamefont
  {M.~H.}\ \bibnamefont {Mikkelsen}},\ }\href@noop {} {\bibfield  {journal}
  {\bibinfo  {journal} {Nature Photonics}\ }\textbf {\bibinfo {volume} {8}},\
  \bibinfo {pages} {835} (\bibinfo {year} {2014})}\BibitemShut {NoStop}%
\bibitem [{\citenamefont {Hoang}\ \emph {et~al.}(2015)\citenamefont {Hoang},
  \citenamefont {Akselrod}, \citenamefont {Argyropoulos}, \citenamefont
  {Huang}, \citenamefont {Smith},\ and\ \citenamefont
  {Mikkelsen}}]{Hoang2015-en}%
  \BibitemOpen
  \bibfield  {author} {\bibinfo {author} {\bibfnamefont {T.~B.}\ \bibnamefont
  {Hoang}}, \bibinfo {author} {\bibfnamefont {G.~M.}\ \bibnamefont {Akselrod}},
  \bibinfo {author} {\bibfnamefont {C.}~\bibnamefont {Argyropoulos}}, \bibinfo
  {author} {\bibfnamefont {J.}~\bibnamefont {Huang}}, \bibinfo {author}
  {\bibfnamefont {D.~R.}\ \bibnamefont {Smith}}, \ and\ \bibinfo {author}
  {\bibfnamefont {M.~H.}\ \bibnamefont {Mikkelsen}},\ }\href@noop {} {\bibfield
   {journal} {\bibinfo  {journal} {Nat. Commun.}\ }\textbf {\bibinfo {volume}
  {6}},\ \bibinfo {pages} {7788} (\bibinfo {year} {2015})}\BibitemShut
  {NoStop}%
\bibitem [{\citenamefont {Karamlou}\ \emph {et~al.}(2018)\citenamefont
  {Karamlou}, \citenamefont {Trusheim},\ and\ \citenamefont
  {Englund}}]{Karamlou2018-cv}%
  \BibitemOpen
  \bibfield  {author} {\bibinfo {author} {\bibfnamefont {A.}~\bibnamefont
  {Karamlou}}, \bibinfo {author} {\bibfnamefont {M.~E.}\ \bibnamefont
  {Trusheim}}, \ and\ \bibinfo {author} {\bibfnamefont {D.}~\bibnamefont
  {Englund}},\ }\href@noop {} {\bibfield  {journal} {\bibinfo  {journal} {Opt.
  Express}\ }\textbf {\bibinfo {volume} {26}},\ \bibinfo {pages} {3341}
  (\bibinfo {year} {2018})}\BibitemShut {NoStop}%
\bibitem [{\citenamefont {Bogdanov}\ \emph {et~al.}(2017)\citenamefont
  {Bogdanov}, \citenamefont {Shalaginov}, \citenamefont {Akimov}, \citenamefont
  {Lagutchev}, \citenamefont {Kapitanova}, \citenamefont {Liu}, \citenamefont
  {Woods}, \citenamefont {Ferrera}, \citenamefont {Belov}, \citenamefont
  {Irudayaraj}, \citenamefont {Boltasseva},\ and\ \citenamefont
  {Shalaev}}]{Bogdanov2017-uu}%
  \BibitemOpen
  \bibfield  {author} {\bibinfo {author} {\bibfnamefont {S.}~\bibnamefont
  {Bogdanov}}, \bibinfo {author} {\bibfnamefont {M.~Y.}\ \bibnamefont
  {Shalaginov}}, \bibinfo {author} {\bibfnamefont {A.}~\bibnamefont {Akimov}},
  \bibinfo {author} {\bibfnamefont {A.~S.}\ \bibnamefont {Lagutchev}}, \bibinfo
  {author} {\bibfnamefont {P.}~\bibnamefont {Kapitanova}}, \bibinfo {author}
  {\bibfnamefont {J.}~\bibnamefont {Liu}}, \bibinfo {author} {\bibfnamefont
  {D.}~\bibnamefont {Woods}}, \bibinfo {author} {\bibfnamefont
  {M.}~\bibnamefont {Ferrera}}, \bibinfo {author} {\bibfnamefont
  {P.}~\bibnamefont {Belov}}, \bibinfo {author} {\bibfnamefont
  {J.}~\bibnamefont {Irudayaraj}}, \bibinfo {author} {\bibfnamefont
  {A.}~\bibnamefont {Boltasseva}}, \ and\ \bibinfo {author} {\bibfnamefont
  {V.~M.}\ \bibnamefont {Shalaev}},\ }\href@noop {} {\bibfield  {journal}
  {\bibinfo  {journal} {Phys. Rev. B Condens. Matter}\ }\textbf {\bibinfo
  {volume} {96}},\ \bibinfo {pages} {035146} (\bibinfo {year}
  {2017})}\BibitemShut {NoStop}%
\bibitem [{\citenamefont {De~Leon}\ \emph {et~al.}(2012)\citenamefont
  {De~Leon}, \citenamefont {Shields}, \citenamefont {Chun}, \citenamefont
  {Englund}, \citenamefont {Akimov}, \citenamefont {Lukin},\ and\ \citenamefont
  {Park}}]{De_Leon2012-ee}%
  \BibitemOpen
  \bibfield  {author} {\bibinfo {author} {\bibfnamefont {N.~P.}\ \bibnamefont
  {De~Leon}}, \bibinfo {author} {\bibfnamefont {B.~J.}\ \bibnamefont
  {Shields}}, \bibinfo {author} {\bibfnamefont {L.~Y.}\ \bibnamefont {Chun}},
  \bibinfo {author} {\bibfnamefont {D.~E.}\ \bibnamefont {Englund}}, \bibinfo
  {author} {\bibfnamefont {A.~V.}\ \bibnamefont {Akimov}}, \bibinfo {author}
  {\bibfnamefont {M.~D.}\ \bibnamefont {Lukin}}, \ and\ \bibinfo {author}
  {\bibfnamefont {H.}~\bibnamefont {Park}},\ }\href@noop {} {\bibfield
  {journal} {\bibinfo  {journal} {Phys. Rev. Lett.}\ }\textbf {\bibinfo
  {volume} {108}},\ \bibinfo {pages} {226803} (\bibinfo {year}
  {2012})}\BibitemShut {NoStop}%
\bibitem [{\citenamefont {Hausmann}\ \emph {et~al.}(2013)\citenamefont
  {Hausmann}, \citenamefont {Shields}, \citenamefont {Quan}, \citenamefont
  {Chu}, \citenamefont {de~Leon}, \citenamefont {Evans}, \citenamefont {Burek},
  \citenamefont {Zibrov}, \citenamefont {Markham}, \citenamefont {Twitchen},
  \citenamefont {Park}, \citenamefont {Lukin},\ and\ \citenamefont {Lon{\v
  c}ar}}]{Hausmann2013-ob}%
  \BibitemOpen
  \bibfield  {author} {\bibinfo {author} {\bibfnamefont {B.~J.~M.}\
  \bibnamefont {Hausmann}}, \bibinfo {author} {\bibfnamefont {B.~J.}\
  \bibnamefont {Shields}}, \bibinfo {author} {\bibfnamefont {Q.}~\bibnamefont
  {Quan}}, \bibinfo {author} {\bibfnamefont {Y.}~\bibnamefont {Chu}}, \bibinfo
  {author} {\bibfnamefont {N.~P.}\ \bibnamefont {de~Leon}}, \bibinfo {author}
  {\bibfnamefont {R.}~\bibnamefont {Evans}}, \bibinfo {author} {\bibfnamefont
  {M.~J.}\ \bibnamefont {Burek}}, \bibinfo {author} {\bibfnamefont {A.~S.}\
  \bibnamefont {Zibrov}}, \bibinfo {author} {\bibfnamefont {M.}~\bibnamefont
  {Markham}}, \bibinfo {author} {\bibfnamefont {D.~J.}\ \bibnamefont
  {Twitchen}}, \bibinfo {author} {\bibfnamefont {H.}~\bibnamefont {Park}},
  \bibinfo {author} {\bibfnamefont {M.~D.}\ \bibnamefont {Lukin}}, \ and\
  \bibinfo {author} {\bibfnamefont {M.}~\bibnamefont {Lon{\v c}ar}},\
  }\href@noop {} {\bibfield  {journal} {\bibinfo  {journal} {Nano Lett.}\
  }\textbf {\bibinfo {volume} {13}},\ \bibinfo {pages} {5791} (\bibinfo {year}
  {2013})}\BibitemShut {NoStop}%
\bibitem [{\citenamefont {Jun}\ \emph {et~al.}(2010)\citenamefont {Jun},
  \citenamefont {Pala},\ and\ \citenamefont {Brongersma}}]{Jun2010-wb}%
  \BibitemOpen
  \bibfield  {author} {\bibinfo {author} {\bibfnamefont {Y.~C.}\ \bibnamefont
  {Jun}}, \bibinfo {author} {\bibfnamefont {R.}~\bibnamefont {Pala}}, \ and\
  \bibinfo {author} {\bibfnamefont {M.~L.}\ \bibnamefont {Brongersma}},\
  }\href@noop {} {\bibfield  {journal} {\bibinfo  {journal} {J. Phys. Chem. C}\
  }\textbf {\bibinfo {volume} {114}},\ \bibinfo {pages} {7269} (\bibinfo {year}
  {2010})}\BibitemShut {NoStop}%
\bibitem [{\citenamefont {Russell}\ \emph {et~al.}(2012)\citenamefont
  {Russell}, \citenamefont {Liu}, \citenamefont {Cui},\ and\ \citenamefont
  {Hu}}]{Russell2012-cn}%
  \BibitemOpen
  \bibfield  {author} {\bibinfo {author} {\bibfnamefont {K.~J.}\ \bibnamefont
  {Russell}}, \bibinfo {author} {\bibfnamefont {T.-L.}\ \bibnamefont {Liu}},
  \bibinfo {author} {\bibfnamefont {S.}~\bibnamefont {Cui}}, \ and\ \bibinfo
  {author} {\bibfnamefont {E.~L.}\ \bibnamefont {Hu}},\ }\href@noop {}
  {\bibfield  {journal} {\bibinfo  {journal} {Nat. Photonics}\ }\textbf
  {\bibinfo {volume} {6}},\ \bibinfo {pages} {459} (\bibinfo {year}
  {2012})}\BibitemShut {NoStop}%
\bibitem [{\citenamefont {Wood}(1935)}]{Wood1935-wq}%
  \BibitemOpen
  \bibfield  {author} {\bibinfo {author} {\bibfnamefont {R.~W.}\ \bibnamefont
  {Wood}},\ }\href@noop {} {\bibfield  {journal} {\bibinfo  {journal} {Phys.
  Rev.}\ }\textbf {\bibinfo {volume} {48}},\ \bibinfo {pages} {928} (\bibinfo
  {year} {1935})}\BibitemShut {NoStop}%
\bibitem [{\citenamefont {Rayleigh}(1907)}]{Rayleigh1907-tp}%
  \BibitemOpen
  \bibfield  {author} {\bibinfo {author} {\bibfnamefont {L.}~\bibnamefont
  {Rayleigh}},\ }\href@noop {} {\bibfield  {journal} {\bibinfo  {journal} {The
  London, Edinburgh, and Dublin Philosophical Magazine and Journal of Science}\
  }\textbf {\bibinfo {volume} {14}},\ \bibinfo {pages} {60} (\bibinfo {year}
  {1907})}\BibitemShut {NoStop}%
\bibitem [{\citenamefont {Ulbricht}\ and\ \citenamefont
  {Loh}(2018)}]{Ulbricht2018-yv}%
  \BibitemOpen
  \bibfield  {author} {\bibinfo {author} {\bibfnamefont {R.}~\bibnamefont
  {Ulbricht}}\ and\ \bibinfo {author} {\bibfnamefont {Z.-H.}\ \bibnamefont
  {Loh}},\ }\href@noop {} {\bibfield  {journal} {\bibinfo  {journal} {Physical
  Review B}\ }\textbf {\bibinfo {volume} {98}} (\bibinfo {year}
  {2018})}\BibitemShut {NoStop}%
\bibitem [{\citenamefont {Tetienne}\ \emph {et~al.}(2012)\citenamefont
  {Tetienne}, \citenamefont {Rondin}, \citenamefont {Spinicelli}, \citenamefont
  {Chipaux}, \citenamefont {Debuisschert}, \citenamefont {Roch},\ and\
  \citenamefont {Jacques}}]{Tetienne2012-ic}%
  \BibitemOpen
  \bibfield  {author} {\bibinfo {author} {\bibfnamefont {J.-P.}\ \bibnamefont
  {Tetienne}}, \bibinfo {author} {\bibfnamefont {L.}~\bibnamefont {Rondin}},
  \bibinfo {author} {\bibfnamefont {P.}~\bibnamefont {Spinicelli}}, \bibinfo
  {author} {\bibfnamefont {M.}~\bibnamefont {Chipaux}}, \bibinfo {author}
  {\bibfnamefont {T.}~\bibnamefont {Debuisschert}}, \bibinfo {author}
  {\bibfnamefont {J.-F.}\ \bibnamefont {Roch}}, \ and\ \bibinfo {author}
  {\bibfnamefont {V.}~\bibnamefont {Jacques}},\ }\href@noop {} {\bibfield
  {journal} {\bibinfo  {journal} {New J. Phys.}\ }\textbf {\bibinfo {volume}
  {14}},\ \bibinfo {pages} {103033} (\bibinfo {year} {2012})}\BibitemShut
  {NoStop}%
\bibitem [{\citenamefont {Robledo}\ \emph {et~al.}(2011)\citenamefont
  {Robledo}, \citenamefont {Bernien}, \citenamefont {van~der Sar},\ and\
  \citenamefont {Hanson}}]{Robledo2011-xg}%
  \BibitemOpen
  \bibfield  {author} {\bibinfo {author} {\bibfnamefont {L.}~\bibnamefont
  {Robledo}}, \bibinfo {author} {\bibfnamefont {H.}~\bibnamefont {Bernien}},
  \bibinfo {author} {\bibfnamefont {T.}~\bibnamefont {van~der Sar}}, \ and\
  \bibinfo {author} {\bibfnamefont {R.}~\bibnamefont {Hanson}},\ }\href@noop {}
  {\bibfield  {journal} {\bibinfo  {journal} {New J. Phys.}\ }\textbf {\bibinfo
  {volume} {13}},\ \bibinfo {pages} {025013} (\bibinfo {year}
  {2011})}\BibitemShut {NoStop}%
\bibitem [{\citenamefont {Wee}\ \emph {et~al.}(2007)\citenamefont {Wee},
  \citenamefont {Tzeng}, \citenamefont {Han}, \citenamefont {Chang},
  \citenamefont {Fann}, \citenamefont {Hsu}, \citenamefont {Chen},\ and\
  \citenamefont {Yu}}]{Wee2007-jo}%
  \BibitemOpen
  \bibfield  {author} {\bibinfo {author} {\bibfnamefont {T.-L.}\ \bibnamefont
  {Wee}}, \bibinfo {author} {\bibfnamefont {Y.-K.}\ \bibnamefont {Tzeng}},
  \bibinfo {author} {\bibfnamefont {C.-C.}\ \bibnamefont {Han}}, \bibinfo
  {author} {\bibfnamefont {H.-C.}\ \bibnamefont {Chang}}, \bibinfo {author}
  {\bibfnamefont {W.}~\bibnamefont {Fann}}, \bibinfo {author} {\bibfnamefont
  {J.-H.}\ \bibnamefont {Hsu}}, \bibinfo {author} {\bibfnamefont {K.-M.}\
  \bibnamefont {Chen}}, \ and\ \bibinfo {author} {\bibfnamefont {Y.-C.}\
  \bibnamefont {Yu}},\ }\href@noop {} {\bibfield  {journal} {\bibinfo
  {journal} {The Journal of Physical Chemistry A}\ }\textbf {\bibinfo {volume}
  {111}},\ \bibinfo {pages} {9379} (\bibinfo {year} {2007})}\BibitemShut
  {NoStop}%
\bibitem [{\citenamefont {Acosta}\ \emph {et~al.}(2009)\citenamefont {Acosta},
  \citenamefont {Bauch}, \citenamefont {Ledbetter}, \citenamefont {Santori},
  \citenamefont {Fu}, \citenamefont {Barclay}, \citenamefont {Beausoleil},
  \citenamefont {Linget}, \citenamefont {Roch}, \citenamefont {Treussart},
  \citenamefont {Chemerisov}, \citenamefont {Gawlik},\ and\ \citenamefont
  {Budker}}]{Acosta2009-te}%
  \BibitemOpen
  \bibfield  {author} {\bibinfo {author} {\bibfnamefont {V.~M.}\ \bibnamefont
  {Acosta}}, \bibinfo {author} {\bibfnamefont {E.}~\bibnamefont {Bauch}},
  \bibinfo {author} {\bibfnamefont {M.~P.}\ \bibnamefont {Ledbetter}}, \bibinfo
  {author} {\bibfnamefont {C.}~\bibnamefont {Santori}}, \bibinfo {author}
  {\bibfnamefont {K.-M.~C.}\ \bibnamefont {Fu}}, \bibinfo {author}
  {\bibfnamefont {P.~E.}\ \bibnamefont {Barclay}}, \bibinfo {author}
  {\bibfnamefont {R.~G.}\ \bibnamefont {Beausoleil}}, \bibinfo {author}
  {\bibfnamefont {H.}~\bibnamefont {Linget}}, \bibinfo {author} {\bibfnamefont
  {J.~F.}\ \bibnamefont {Roch}}, \bibinfo {author} {\bibfnamefont
  {F.}~\bibnamefont {Treussart}}, \bibinfo {author} {\bibfnamefont
  {S.}~\bibnamefont {Chemerisov}}, \bibinfo {author} {\bibfnamefont
  {W.}~\bibnamefont {Gawlik}}, \ and\ \bibinfo {author} {\bibfnamefont
  {D.}~\bibnamefont {Budker}},\ }\href@noop {} {\bibfield  {journal} {\bibinfo
  {journal} {Phys. Rev. B Condens. Matter}\ }\textbf {\bibinfo {volume} {80}},\
  \bibinfo {pages} {115202} (\bibinfo {year} {2009})}\BibitemShut {NoStop}%
\bibitem [{\citenamefont {Meng}\ \emph {et~al.}(2010)\citenamefont {Meng},
  \citenamefont {Lang}, \citenamefont {Wessel}, \citenamefont {Assaad},\ and\
  \citenamefont {Muramatsu}}]{Meng2010-qp}%
  \BibitemOpen
  \bibfield  {author} {\bibinfo {author} {\bibfnamefont {Z.~Y.}\ \bibnamefont
  {Meng}}, \bibinfo {author} {\bibfnamefont {T.~C.}\ \bibnamefont {Lang}},
  \bibinfo {author} {\bibfnamefont {S.}~\bibnamefont {Wessel}}, \bibinfo
  {author} {\bibfnamefont {F.~F.}\ \bibnamefont {Assaad}}, \ and\ \bibinfo
  {author} {\bibfnamefont {A.}~\bibnamefont {Muramatsu}},\ }\href@noop {}
  {\bibfield  {journal} {\bibinfo  {journal} {Nature}\ }\textbf {\bibinfo
  {volume} {464}},\ \bibinfo {pages} {847} (\bibinfo {year}
  {2010})}\BibitemShut {NoStop}%
\bibitem [{\citenamefont {Takagi}\ \emph {et~al.}(2019)\citenamefont {Takagi},
  \citenamefont {Takayama}, \citenamefont {Jackeli}, \citenamefont
  {Khaliullin},\ and\ \citenamefont {Nagler}}]{Takagi2019-hl}%
  \BibitemOpen
  \bibfield  {author} {\bibinfo {author} {\bibfnamefont {H.}~\bibnamefont
  {Takagi}}, \bibinfo {author} {\bibfnamefont {T.}~\bibnamefont {Takayama}},
  \bibinfo {author} {\bibfnamefont {G.}~\bibnamefont {Jackeli}}, \bibinfo
  {author} {\bibfnamefont {G.}~\bibnamefont {Khaliullin}}, \ and\ \bibinfo
  {author} {\bibfnamefont {S.~E.}\ \bibnamefont {Nagler}},\ }\href@noop {}
  {\bibfield  {journal} {\bibinfo  {journal} {Nature Reviews Physics}\ }\textbf
  {\bibinfo {volume} {1}},\ \bibinfo {pages} {264} (\bibinfo {year}
  {2019})}\BibitemShut {NoStop}%
\bibitem [{\citenamefont {Cappellaro}\ and\ \citenamefont
  {Lukin}(2009)}]{Cappellaro2009-os}%
  \BibitemOpen
  \bibfield  {author} {\bibinfo {author} {\bibfnamefont {P.}~\bibnamefont
  {Cappellaro}}\ and\ \bibinfo {author} {\bibfnamefont {M.~D.}\ \bibnamefont
  {Lukin}},\ }\href@noop {} {\bibfield  {journal} {\bibinfo  {journal} {Phys.
  Rev. A}\ }\textbf {\bibinfo {volume} {80}},\ \bibinfo {pages} {032311}
  (\bibinfo {year} {2009})}\BibitemShut {NoStop}%
\bibitem [{\citenamefont {Choi}\ \emph {et~al.}(2017)\citenamefont {Choi},
  \citenamefont {Yao},\ and\ \citenamefont {Lukin}}]{Choi2017-lu}%
  \BibitemOpen
  \bibfield  {author} {\bibinfo {author} {\bibfnamefont {S.}~\bibnamefont
  {Choi}}, \bibinfo {author} {\bibfnamefont {N.~Y.}\ \bibnamefont {Yao}}, \
  and\ \bibinfo {author} {\bibfnamefont {M.~D.}\ \bibnamefont {Lukin}},\
  }\href@noop {} {\bibfield  {journal} {\bibinfo  {journal} {arXiv}\ }
  (\bibinfo {year} {2017})},\ \Eprint {http://arxiv.org/abs/1801.00042}
  {arXiv:1801.00042 [quant-ph]} \BibitemShut {NoStop}%
\end{thebibliography}%
\end{document}


\title{Supplementary Information for Absorption-Based Diamond Spin Microscopy on a Plasmonic Quantum Metasurface}
\author{Laura Kim}
\affiliation{Research Laboratory of Electronics, MIT, Cambridge, MA 02139, USA}

\author{Hyeongrak Choi}
\affiliation{Research Laboratory of Electronics, MIT, Cambridge, MA 02139, USA}
\affiliation{Department of Electrical Engineering and Computer Science, MIT, Cambridge, MA 02139, USA}

\author{Matthew E. Trusheim}
\affiliation{Research Laboratory of Electronics, MIT, Cambridge, MA 02139, USA}
\affiliation{U.S. Army Research Laboratory, Sensors and Electron Devices Directorate, Adelphi, Maryland 20783, USA}

\author{Dirk R. Englund}
\affiliation{Research Laboratory of Electronics, MIT, Cambridge, MA 02139, USA}
\affiliation{Department of Electrical Engineering and Computer Science, MIT, Cambridge, MA 02139, USA}
\date{\today}

\maketitle

\section{Section 1: Metal-diamond metallodielectric grating structures}
\begin{figure}[hbt!]
    \centering
    \includegraphics[width=0.65\columnwidth]{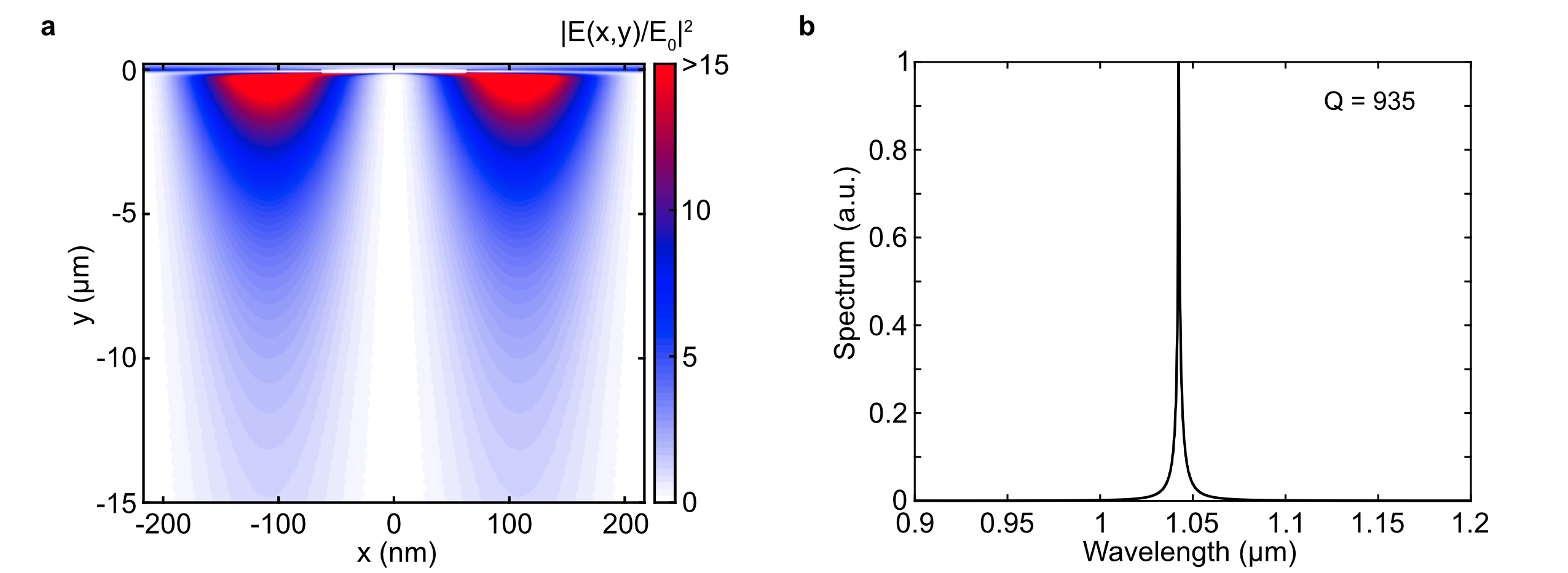}
    \caption{Silver-diamond metallodielectric grating (a) Total electric field intensity with an extended y-range (b) The spectrum obtained by Fourier transform of the time decay of the field at 287.8 THz shows Q = 935.618.}
    \label{fig:Sfig1}
\end{figure}

The SPP-RWA hybrid mode of the PQSM creates a vertically extended field down to about 5 \textmu m in depth (Fig.\,\ref{fig:Sfig1}a), and exhibits a quality factor of 935 (Fig.\,\ref{fig:Sfig1}b).

\begin{figure}[hbt!]
    \centering
    \includegraphics[width=0.7\textwidth]{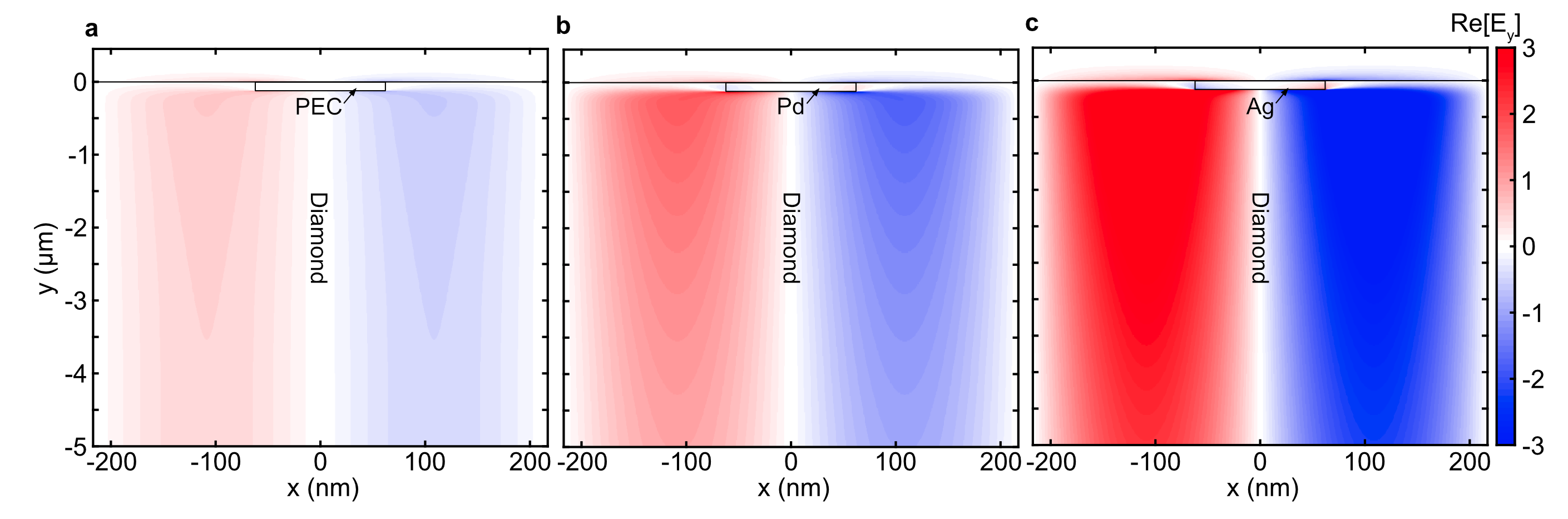}
    \caption{Field distribution, Re($E_y$), of a grating structure made with diamond - (a) PEC (b) palladium and (c) silver with $p$ = 434 nm, $w$ = 125 nm, and $t$ = 125nm.}
    \label{fig:Sfig2}
\end{figure}

The dispersion for SPPs under the Bragg grating coupling condition (i.e., surface plasmon polariton-Bloch wave (SPP-BW) modes) is given by Eq.\,\ref{eq:Seq1} \cite{Barnes2004-hl}.
\begin{equation}\label{eq:Seq1}
    \text{Re}\Big[\frac{\omega}{c}\sqrt{\frac{\epsilon_\text{m}\epsilon_\text{d}}{\epsilon_\text{m}+\epsilon_\text{d}}}\Big]=\Big|k_x+m\frac{2\pi}{p}\Big|
\end{equation}
where $\epsilon_\text{m/d}$ is the permittivity of metallic/dielectric material of the metallodielectric grating structure and $m$ is the integer denoting specific SPP modes \cite{Gao2009-fk,Barnes2004-hl}.
While RWA modes are independent of metal properties, the SPP-RWA hybrid becomes highly dependent on the properties of the plasmonic material. The SPPs play a role in creating a large field enhancement in the metallodielectric grating structure, and a dramatic reduction in the electric field intensity is observed when silver is replaced with a weak plasmonic material, palladium (Fig.\,\ref{fig:Sfig2}b). It is evident that the RWA is responsible for creating a vertically extended electric field as it is maintained even when silver is replaced with a perfect electric conductor (PEC), which cannot support SPPs (Fig.\,\ref{fig:Sfig2}). 

\begin{figure}[hbt!]
    \centering
    \includegraphics[width=0.4\columnwidth]{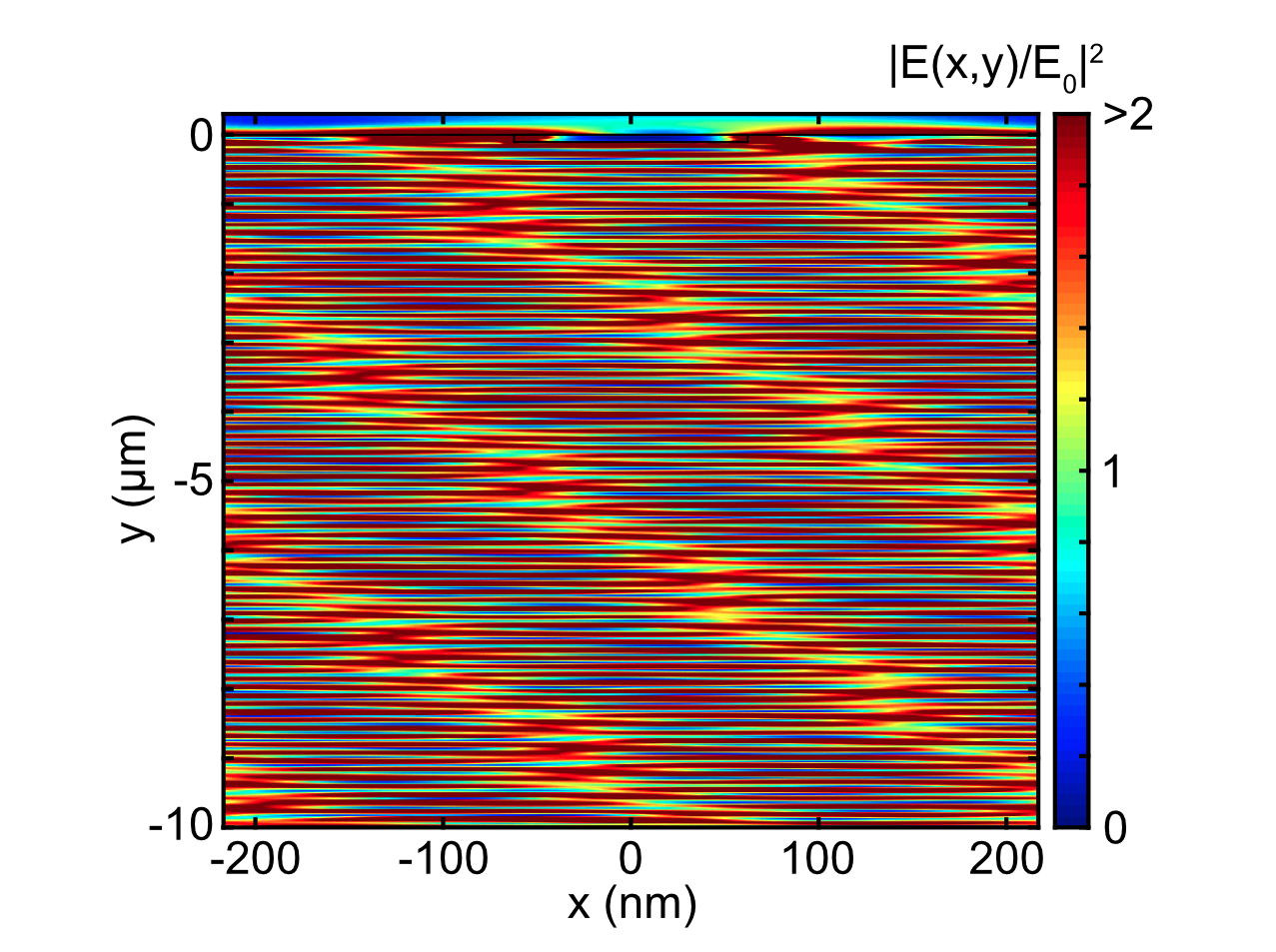}
    \caption{Total electric field intensity of a Ag-diamond grating structure with $p$ = 434 nm, $w$ = 125 nm, and $t$ = 125 nm under TM-polarized illumination at 532 nm from the backside with an off-normal incidence ($\theta_i$ = 2.9$^\circ$).}
    \label{fig:Sfig3}
\end{figure}

Without changing geometric parameters of the PQSM, we can couple in the green laser excitation, which is necessary to populate the singlet state, by inducing a grating resonance at 532 nm via a off-normal incidence (i.e., compensating for the momentum mismatch, $k_x$). For example, for the given structure geometry studied in this work, TM-polarized illumination at 532 nm from the backside with $\theta_i$ = 2.9$^\circ$ satisfies the condition for the RWA mode for $m$ = 2. Figure\,\ref{fig:Sfig3} shows the electric field intensity profile, and the corresponding spatially averaged electric field intensity, $\braket{|E/E_0|^2}$, is 1.98.

\section{Section 2: Rate equations for the 8-level scheme}
The local density of each sublevel, $n_{\ket{i}}$, is calculated based on the coupled rate equations (Eq.\,\ref{eq:n1} - Eq.\,\ref{eq:n8}) under the assumption of spin-conserving optical transition. We impose a number conservation constraint (i.e., $\sum\limits_{i} n_{\ket{i}}=n_\text{NV}$, where $n_\text{NV}$ is the total NV density).

\begin{subequations}
\begin{align}
    \frac{dn_{\ket{1}}}{dt} &= -(W_\text{pump}+W_\text{MW})n_{\ket{1}}+W_\text{MW}n_{\ket{2}}+k_{31}n_{\ket{3}}+k_{61}n_{\ket{6}}+k_{71}n_{\ket{7}}\label{eq:n1}\\
    \frac{dn_{\ket{2}}}{dt} &= W_\text{MW}n_{\ket{1}}-(W_\text{pump}+W_\text{MW})n_{\ket{2}}+k_{42}n_{\ket{4}}+k_{62}n_{\ket{6}}+k_{72}n_{\ket{7}}\label{eq:n2}\\
    \frac{dn_{\ket{3}}}{dt} &= W_\text{pump}n_{\ket{1}}-(k_{31}+k_{35}+k_{38})n_{\ket{3}}\label{eq:n3}\\
    \frac{dn_{\ket{4}}}{dt} &= W_\text{pump}n_{\ket{2}}-(k_{42}+k_{45}+k_{48})n_{\ket{4}}\label{eq:n4}\\
    \frac{dn_{\ket{5}}}{dt} &= k_{35}n_{\ket{3}}+k_{45}n_{\ket{4}}-(\gamma_\text{nr}+F_p\gamma_\text{r}+\gamma_\text{quenching}+W_\text{probe})n_{\ket{5}}+W_\text{probe}n_{\ket{6}}\label{eq:5}\\
    \frac{dn_{\ket{6}}}{dt} &= (\gamma_\text{nr}+F_p\gamma_\text{r}+\gamma_\text{quenching}+W_\text{probe})n_{\ket{5}}-(W_\text{probe}+k_{61}+k_{62})n_{\ket{6}}\label{eq:n6}\\
    \frac{dn_{\ket{7}}}{dt} &= -(\Gamma_{\text{NV}^0}+k_{72}+k_{71})n_{\ket{7}}+W_{\text{NV}^0}n_{\ket{8}}\label{eq:n7}\\
    \frac{dn_{\ket{8}}}{dt} &= k_{38}n_{\ket{3}}+k_{48}n_{\ket{4}}+\Gamma_{\text{NV}^0}n_{\ket{7}}-W_{\text{NV}^0}n_{\ket{8}}\label{eq:n8}
\end{align}
\end{subequations}

$k_{ij}$ is the transition rate constant from state $\ket{i}$ to state $\ket{j}$. $\Gamma_{\text{NV}^0}$ is the decay rate from $\ket{7}$ to $\ket{8}$, and $\Gamma=\gamma_\text{nr}+F_p\gamma_\text{r}+\gamma_\text{quenching}$ is the decay rate from $\ket{5}$ to $\ket{6}$, where $F_p$ is the Purcell factor, $\gamma_\text{quenching}$ is the quenching rate, and $\gamma_\text{nr}$ and $\gamma_\text{r}$ are the non-radiative and radiative decay rates of state $\ket{5}$,  respectively. $W_\text{pump/probe}$ is the optical excitation rate of the triplet/singlet transition given by $\frac{\sigma_\text{t/s}I_\text{t/s}}{\hbar\omega_\text{t/s}}$, where $\sigma$ is the absorption cross sectional area, $I$ is the laser intensity, and $\omega$ is the angular frequency of the corresponding transition. The PQSM enhances the $W_\text{pump/probe}$ by a factor of $\sim\braket{|E/E_0|^2}$ at $\lambda$ = 532 nm/1042 nm. $W_\text{MW}$ is the microwave transition rate approximated as $\Omega_R^2 T_2^*/2$, where $\Omega_R$ is the Rabi angular frequency and $T_2^*$ is the electron spin dephasing time \cite{Dumeige2013-wf}. All of the relevant parameters are listed in Table 1 of the main manuscript. 

\section{Section 3: Derivation for SNR of the PQSM}
The intrinsic rate of absorption can be written in terms of the intrinsic absorption cross section of the singlet state transition, $\sigma_\text{s}$ (Eq.\,\ref{eq:Seq10}).
\begin{equation}\label{eq:Seq10}
    \Gamma_0=\frac{\sigma_\text{s}I_s}{\hbar\omega_0}
\end{equation}
The plasmonic structures explored in this work enhance the rate of absorption of an emitter at the position ($x$,$y$,$z$) by a factor of $|E(x,y)/E_0|^2$, where $E_0$ is the electric field in a homogeneous environment and $E(x,y)$ is the electric field induced by the SPP-RWA hybrid mode of the PQSM, invariant in z-direction. The fractional change in IR intensity due to NV absorption, $I_\text{NV}(I_t,\Omega_R)$, for a given pixel size of $L^2$ illuminated with $I_s$ is defined as Eq.\,\ref{eq:Seq12}.
\begin{equation}\label{eq:Seq12}
\begin{split}
    I_\text{NV}(I_t,\Omega_R)&=\frac{I_\text{out}(0,0)-I_\text{out}(I_t,\Omega_R)}{I_\text{out}(0,0)}\\
    &=\frac{I_s}{I_\text{out}(0,0)}\frac{\sigma_\text{s}N_\text{NV}(I_t,\Omega_R)}{L^2}\frac{\int_0^{d_\text{NV}}\int_{-p/2}^{p/2} |\frac{E(x,y)}{E_0}|^2dx dy}{\int_0^{d_\text{NV}}\int_{-p/2}^{p/2}dx dy}
\end{split}
\end{equation}
where $N_\text{NV}(I_t,\Omega_R)$ is the total net NV population in the ground singlet state for a given $V_\text{pixel}$, which is given by $(n_{\ket{6}}-n_{\ket{5}})V_\text{pixel}$. The net population of the singlet ground state (i.e., $n_{\ket{6}}-n_{\ket{5}}$) is used to account for stimulated emission. The SNR is proportional to $I_\text{NV}(I_t,\Omega_R)-I_\text{NV}(I_t,0)$, which is expressed as Eq.\,\ref{eq:Seq13} in the limit of low contrast (i.e., $I_\text{out}(I_t,\Omega_R) \approx I_\text{out}(I_t,0) \approx I_\text{out}(0,0)$). 

\begin{equation}\label{eq:Seq13}
\begin{split}
    I_\text{NV}(I_t,\Omega_R)-I_\text{NV}(I_t,0)&=\frac{\sigma_\text{s}[(n_{\ket{6}}-n_{\ket{5}})_{\Omega_R}-(n_{\ket{6}}-n_{\ket{5}})_0]V_\text{pixel}}{L^2 R_0}\frac{\int_0^{d_\text{NV}}\int_{-p/2}^{p/2} |\frac{E(x,y)}{E_0}|^2dx dy}{\int_0^{d_\text{NV}}\int_{-p/2}^{p/2}dx dy}\\
    &\propto \braket{|E/E_0|^2}V_\text{pixel}n_\text{NV}
\end{split}
\end{equation}
where $R_0$ is the reflection of the PQSM. Thus, the SNR of the pixelated plasmonic imaging surface under the assumption of the shot noise limit is given by Eq.\,\ref{eq:Seq14}.

\begin{equation}\label{eq:Seq14}
    \text{SNR}=\sqrt{\frac{I_s\Delta t_\text{mea}L^2}{2R_0\hbar\omega_0}}\frac{\sigma_\text{s}[(n_{\ket{6}}-n_{\ket{5}})_{\Omega_R}-(n_{\ket{6}}-n_{\ket{5}})_0]}{p}\int_0^{d_\text{NV}}\int_{-p/2}^{p/2} |\frac{E(x,y)}{E_0}|^2dx dy
\end{equation}

\begin{figure}[hbt!]
    \centering
    \includegraphics[width=0.4\columnwidth]{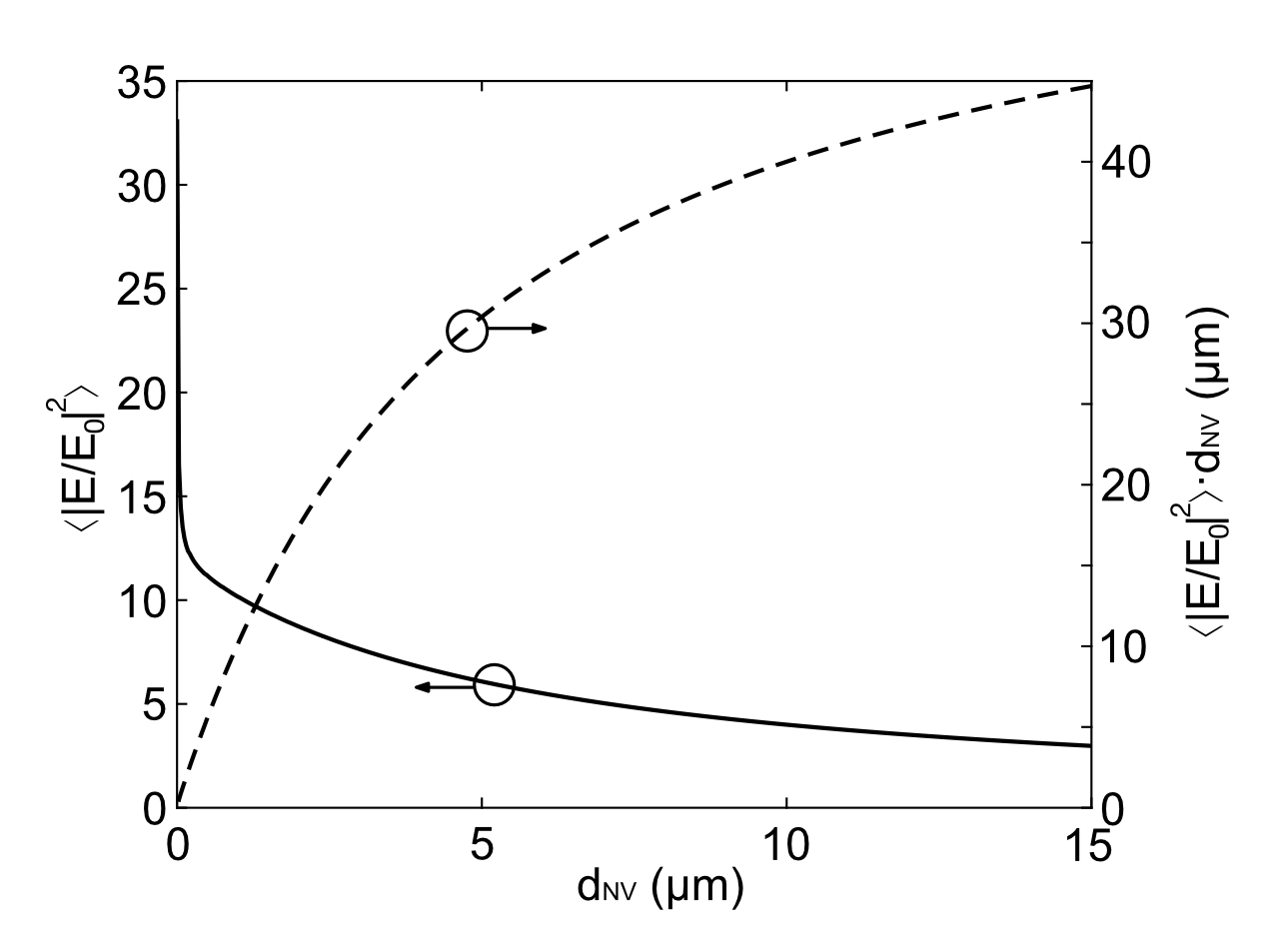}
    \caption{Spatially averaged electric field intensity, $\braket{|E/E_0|^2}$ (left), and the figure of merit, $\braket{|E/E_0|^2}d_\text{NV}$ (right), of the PQSM.}
    \label{fig:Sfig4}
\end{figure}

The performance of an ensemble-based sensor scales with $\braket{|E/E_0|^2}V_\text{pixel}$. Although the electric field is highly concentrated near the metal surface as shown in the plotted $\braket{|E/E_0|^2}$ (Fig. S4), the figure of merit, $\braket{|E/E_0|^2}V_\text{pixel}$, increases with $d_\text{NV}$ (or $V_\text{pixel}$ for a given pixel size of $L^2$).

\section{Section 4: Optimizing homodyne detection condition}
To find the optimal operating condition for homodyne detection, a combination of $R$ and $\Delta\phi_\text{LO}$ that maximizes SNR/$\sqrt{L^2}$ (Eq. S14) is found numerically for given $I_s$ and $I_t$,  while $I_\text{out}(I_t,\Omega_R)$ is replaced by $I_\text{out}(I_t,\Omega_R,R,\Delta\phi_\text{LO})$ (Eq. 6 in the main manuscript). Figure\,\ref{fig:Sfig6} shows the contour plot of the area-normalized SNR as a function of $\Delta\phi_\text{LO}$ and $R$ for given $I_s$, $I_t$, and $\Delta t_\text{mea}$. The optimal condition for homodyne detection is found at $\Delta\phi_\text{LO}$ = 1.28$\pi$ and $R$ = 0.87.

\begin{figure}[hbt!]
    \centering
    \includegraphics[width=0.35\columnwidth]{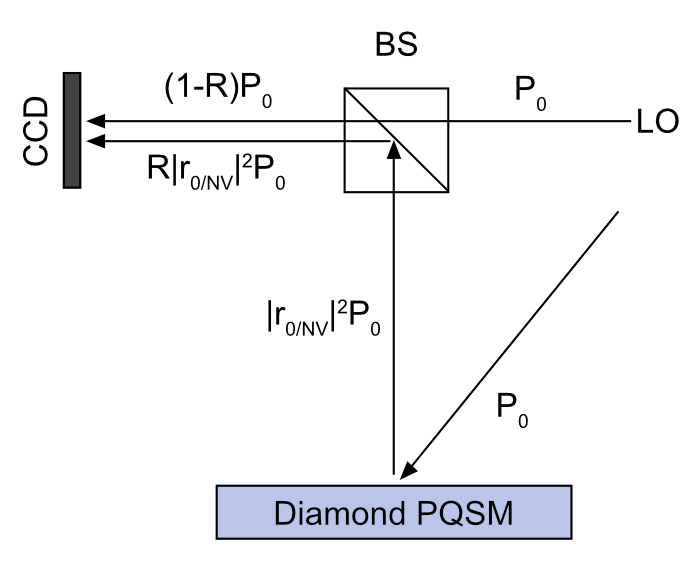}
    \caption{Homodyne detection. $P_0$ is the incident power at 1042 nm given by $I_sL^2$. The metasurface signal, $|r_{0/\text{NV}}|^2 P_0$, and the local oscillator signal, $P_0$, are the input signals to a beam splitter (BS) with a power splitting ratio of $R$, and the interfered signal is detected by a CCD camera.}
    \label{fig:Sfig5}
\end{figure}

\begin{figure}[hbt!]
    \centering
    \includegraphics[width=0.4\columnwidth]{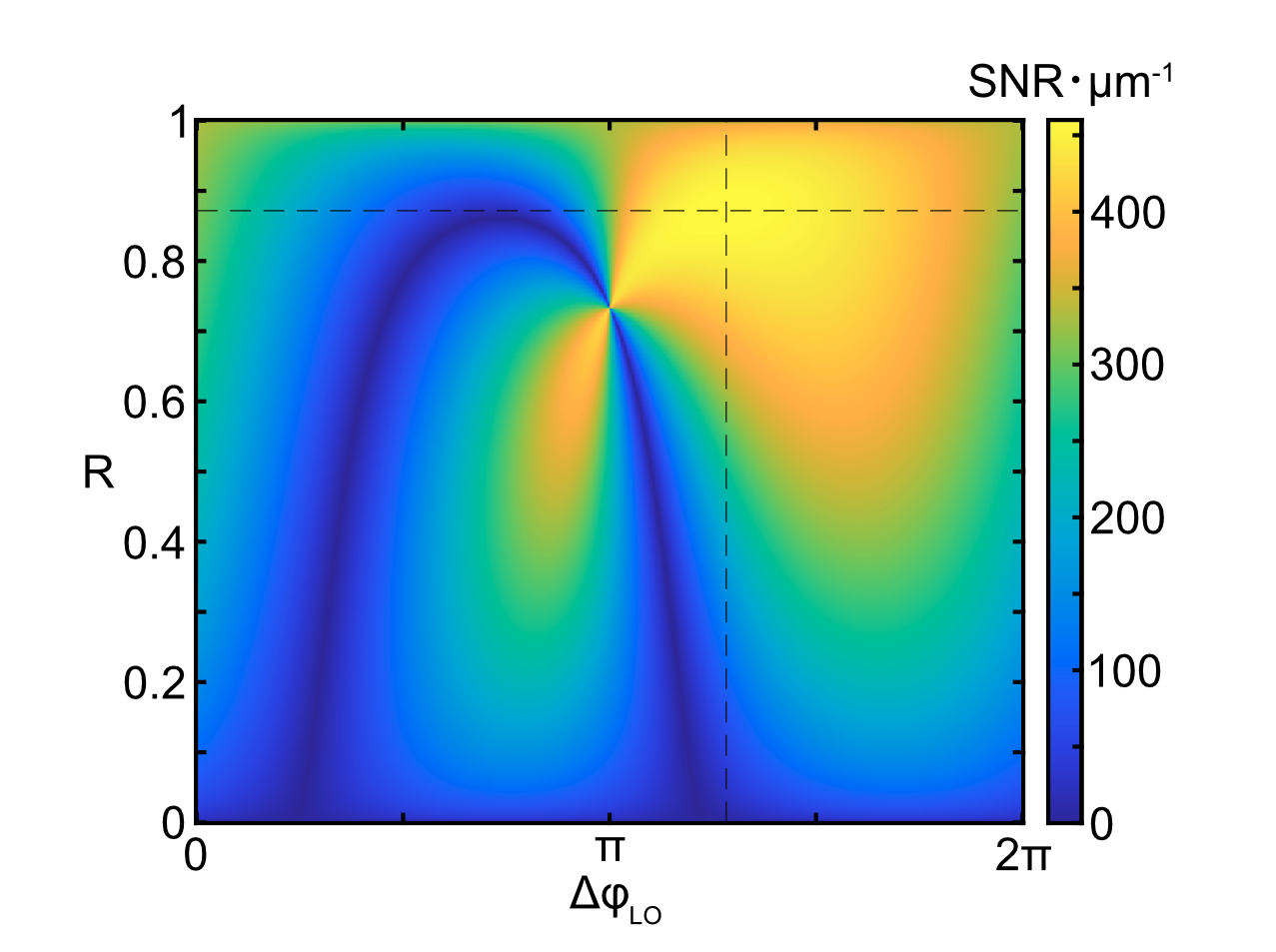}
    \caption{Contour plot of SNR normalized by the pixel area of $L^2$ as a function of $R$ and $\Delta\phi_\text{LO}$ for given $I_s$ = 1 mW/\textmu m$^2$, $I_t$ = 0.1 mW/\textmu m$^2$, and $\Delta t_\text{mea}$ = 10 \textmu s. The maximum is found at $\Delta \phi_\text{LO}$ = 1.28$\pi$ and $R$ = 0.87, indicated by the crossover of the two dotted lines.}
    \label{fig:Sfig6}
\end{figure}

\section{Section 5: Optimal readout condition for pulsed measurements}
\begin{figure}[hbt!]
    \centering
    \includegraphics[width=0.4\columnwidth]{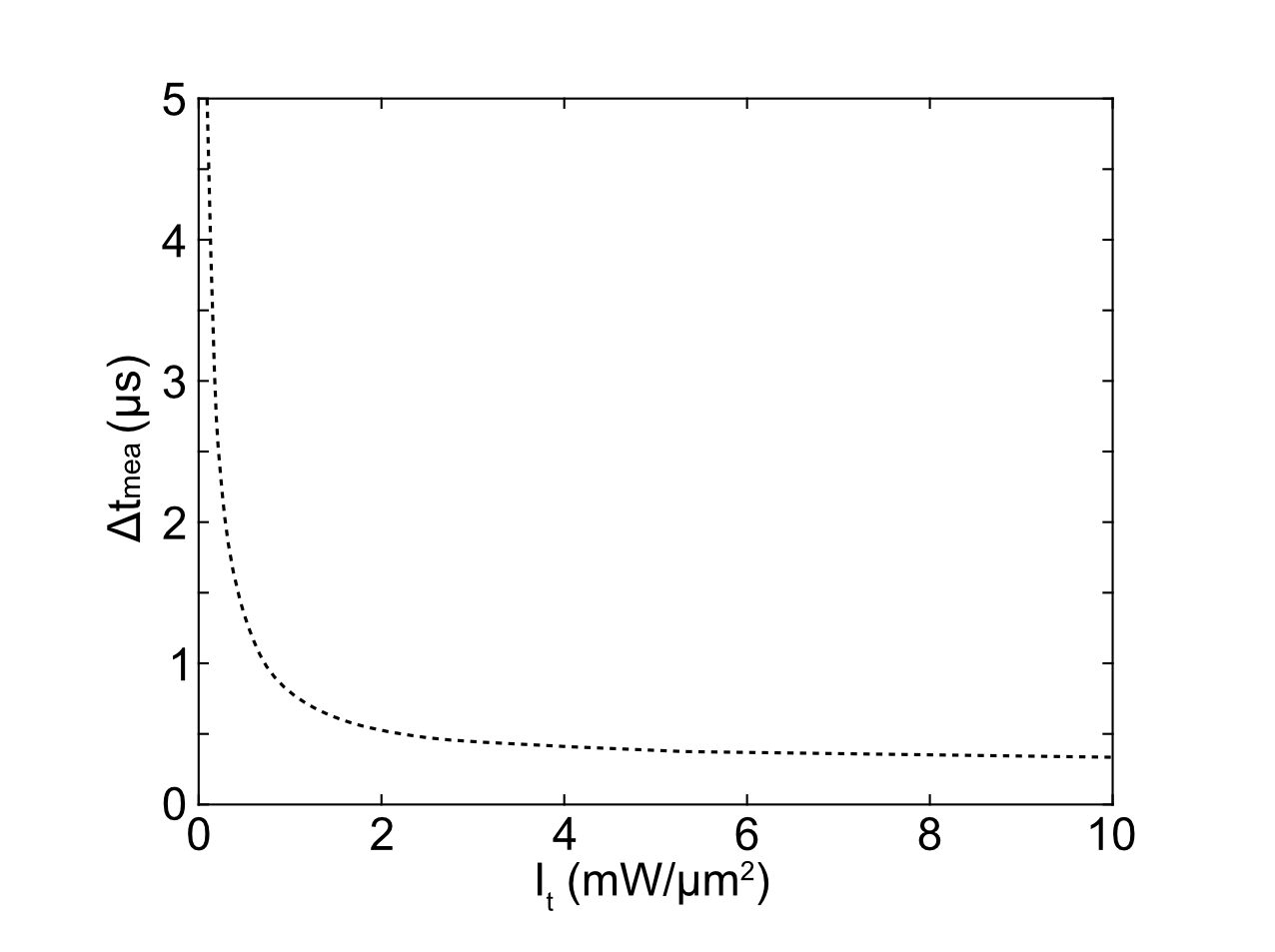}
    \caption{Optimal readout time for pulsed measurements for a given $d_\text{NV}$ = 5 \textmu m.}
    \label{fig:Sfig7}
\end{figure}

As shown in Fig. 4a of the main manuscript, in pulsed measurements, the system achieves maximum spin contrast within the first 1 \textmu s of readout and finally loses polarization with increasing readout time, limited by the lifetime of the ground singlet state. Thus, there exists an optimal readout time that gives the maximum time-averaged signal. For given $I_t$ and $I_s$, the optimal readout time, $\Delta t_\text{mea}$, that maximizes the quantity of Eq.\,\ref{eq:Seq15} is found, and the result is plotted in Fig.\,\ref{fig:Sfig7}.
\begin{equation}\label{eq:Seq15}
    \frac{\int_0^{t_\text{mea}}[(n_{\ket{6}}(t))_{\Omega_R}-(n_{\ket{6}}(t))_0]dt}{\int_0^{t_\text{mea}}dt}
\end{equation}


%